# Bilayer Lithium Niobate Acoustic Resonators for Spurious-Free Wideband Operation at 6 GHz

Florian Hartmann[1*], Silvan Stettler[1], Luis Guillermo Villanueva[1]

[1] Advanced Nanoelectromechanical Systems Laboratory, École Polytechnique Fédérale de Lausanne (EPFL), 1015 Lausanne, Switzerland

[*]Corresponding author :

- E-mail : florian.hartmann@epfl.ch

Contributing authors :

silvan.stettler@epfl.ch, guillermo.villanueva@epfl.ch

# Abstract

The evolution of wireless communication standards toward higher frequencies and wider bandwidths places increasing demands on acoustic filter technologies. Conventional resonators face scaling limitations above 5 GHz, where lithographic constraints, reduced electromechanical coupling coefficient ($k^2_{eff}$), and spurious mode excitation hinder practical filter implementation. Here, we introduce a bilayer X-cut lithium niobate ($LiNbO_3$) thickness-shear bulk acoustic resonator architecture that leverages symmetry engineering as a new design degree of freedom. By vertically stacking two piezoelectric thin films with a tailored bonding angle, the second-order thickness-shear mode ($SH_{2, slow}$) is selectively excited. Furthermore, the fast thickness-shear modes inherent to X-cut $LiNbO_3$ are suppressed, yielding a spurious-free in-band response. Devices operating around 6 GHz demonstrate a $k^2_{eff}$ of approximately 35%, in close agreement with finite-element simulations and confirming the symmetry-driven mode selection mechanism. Proof-of-concept ladder filters demonstrate a fractional bandwidth of 18.6%, highlighting the wideband capabilities of the proposed architecture. The proposed configuration enables thickness-defined frequency scaling while maintaining high coupling and spectral purity, offering a promising platform for wideband radiofrequency filters in next-generation wireless systems.

# Introduction

With the deployment of fifth generation (5G) wireless networks and the ongoing development of future wireless standards, the general trend has been toward higher operating frequencies and wider bandwidths in response to spectrum congestion and the demand for higher data rates [1]. In licensed spectrum, 5G operation in the sub-6 GHz frequency range (FR1) has enabled increased capacity and throughput in both dense urban and rural environments. Key FR1 bands such as n77 (3.3-4.2 GHz), n78 (3.3-3.8 GHz) and n79 (4.4-5 GHz) extend beyond the frequency range traditionally used by 4G Long Term Evolution (LTE) systems, with increased carrier frequencies accompanied by significantly wider channel bandwidths. In parallel, the recent allocation of unlicensed spectrum in the 6 GHz band for Wi-Fi 6E has introduced 1200 MHz of additional bandwidth [2]. These expanded spectral resources require RF front-end systems capable of handling higher frequencies and wider bandwidths. In particular, acoustic filter technologies remain a critical bottleneck, as their performance limits the achievable frequency range and bandwidth of RF front-end filtering solutions [3].

Indeed, conventional surface acoustic wave (SAW) and bulk acoustic wave (BAW) technologies have successfully met the requirements of previous generation wireless systems and remain suitable for part of the 5G spectrum. However, further development toward higher frequencies and wider bandwidths presents different challenges. For SAW devices, increasing the operating frequency requires a corresponding reduction of the acoustic wavelength and increasingly small electrode dimensions, imposing stringent fabrication constraints at the limits of lithography. Developments in incredible high performance SAW (IHP-SAW) extend the performance of SAW devices by improving acoustic energy confinement, quality factor and electromechanical coupling ($k^2_{eff}$) [4–7]. In contrast, BAW devices are not constrained by lateral scaling since the frequency is primarily thickness-defined. Aluminium nitride (AlN) and aluminium scandium nitride (AlScN) remain among the materials of choice for commercial BAW filters, with AlScN substantially increasing $k^2_{eff}$ in comparison with AlN [8–13]. Further advances based on epitaxial AlN[14] have extended BAW operation toward higher frequencies, while periodically poled piezoelectric film (P3F) architectures based on AlN and AlScN enable the excitation of higher-order thickness modes and operation in X and Ku bands [15–17]. Despite these advances, simultaneously achieving very high frequencies and large $k^2_{eff}$ required for the wide fractional bandwidths of emerging standards remains challenging.

The large intrinsic electromechanical coupling of $LiNbO_3$ holds strong potential for simultaneously achieving both requirements. Thin films of $LiNbO_3$ integrated on high acoustic velocity substrates such as SiC enable operation above 3 GHz with enhanced electromechanical coupling and improved quality factor [18–20]. Further improvements on acoustic energy confinement can be achieved using suspended thin-film structures, enabling improved $k^2_{eff}$. Using different electrode configurations, suspended $LiNbO_3$ resonators have demonstrated large coupling ($k^2_{eff}$ > 30%, see definition in Materials and Methods) for symmetric[21–24], shear horizontal modes[25–28] and shear bulk modes[29]. However, their resonance frequency remains strongly dependent on lateral dimensions, limiting aggressive frequency scaling. To alleviate this lateral scaling constraint, higher-order Lamb modes, such as the $A_1$ mode, increasingly exploit the film thickness to achieve high operating frequencies, while maintaining large $k^2_{eff}$ [30,31]. In the large pitch regime, laterally excited bulk acoustic resonators (XBARs) further exploit thickness-dominated operation and have successfully demonstrated large $k^2_{eff}$ at high frequencies [32–34]. Alternatively, high frequency BAW modes can be excited with even larger $k^2_{eff}$ by applying an electric field along the thickness of the $LiNbO_3$ film, as in free-standing thin-film bulk acoustic resonators (FBARs). Several crystalline orientations of $LiNbO_3$ have been demonstrated to be suitable for thickness field excitation of longitudinal (e.g. YX36° [35]) or shear (Y-cut [36,37], YX163° [38], X-cut [39–41]) BAW resonances. In particular, X-cut $LiNbO_3$ supports a slow shear fundamental ($SH_{1,\, slow}$) mode with exceptionally strong electromechanical coupling, which has enabled resonators with $k^2_{eff}$ values around 40%. These values are among the highest reported for acoustic resonator technologies. However, operation of the fundamental mode in frequency bands such as Wi-Fi 6E requires frequencies approaching 6-7 GHz, necessitating aggressive thickness scaling that can affect $k^2_{eff}$, quality factors, and power handling. In this regard, P3F architectures exploiting higher-order thickness shear modes provide an attractive approach to further increase the frequency while avoiding excessive reduction of the piezoelectric thickness and degradation of the performance.

Indeed, P3F $LiNbO_3$ multilayers provide a route to efficiently excite higher-order modes through engineered piezoelectric symmetry. While these architectures have enabled higher-order Lamb-mode operation into the GHz regime [42–44], experimental demonstrations of vertically excited thickness-shear modes in bilayer $LiNbO_3$ have remained at substantially lower frequencies [45,46].

In this work, we present a bilayer X-cut $LiNbO_3$ thickness-shear bulk acoustic resonator for wideband RF filtering applications in the 6-7 GHz range. The proposed bilayer architecture combines +X-cut and -X-cut $LiNbO_3$ thin films with a carefully selected bonding angle to selectively excite the $SH_{2,slow}$ mode, enabling frequency scaling without aggressive thickness reduction while maintaining large $k^2_{eff}$. In addition, the bilayer symmetry suppresses the intrinsic fast shear modes present in conventional single-layer X-cut resonators, resulting in a clean spectral response suitable for filter implementation. The resulting bilayer resonators operate around 6 GHz with $k^2_{eff}$ approaching 35%. The critical role of the stacked layers symmetry and bonding angle are assessed through simulations and experimental measurements. Finally, proof-of-concept ladder-type filters demonstrate the potential of the proposed architecture for wideband RF filtering applications.

# Results

## Bilayer thickness-shear bulk acoustic resonator

Suspended $LiNbO_3$ thickness-shear BAW resonators consist of a suspended $LiNbO_3$ membrane patterned with metallic interdigital transducers (IDTs) on the top surface and a floating electrode on the bottom surface (**Fig. 1a**). This electrode configuration generates a predominantly vertical electric field with alternating polarity, enabling the excitation of thickness-shear modes. The thickness of the piezoelectric stack sets (to leading order) the resonance frequency, while the IDT pitch, the aperture and the trench are used to tune the static capacitance, electrical resistance and the prominence of spurious Lamb modes. We first consider the conventional configuration, in which the piezoelectric stack consists of a single X-cut $LiNbO_3$ film (**Fig. 1b**). In **Fig. 1c**, we show the simulated admittance response of the single layer X-cut $LiNbO_3$ configuration. The simulated admittance response exhibits a strongly coupled $SH_{1, slow}$ resonance. However, the additional $SH_{1, fast}$ mode produces an in-band spurious ripple complicating filter implementation (see **Fig. S1a** in Supplementary Information).

To overcome this limitation, alternative approaches based on multilayer piezoelectric structures have recently been explored, providing additional degrees of freedom for acoustic mode engineering. **Fig. 1a** and **1d** present the bilayer architecture proposed in this work, consisting of stacked X-cut $LiNbO_3$ and -X-cut $LiNbO_3$ layers with the same electrode configuration as in the single layer device (**Fig. 1d**). Depending on the bonding angle and layer thicknesses, the bilayer architecture can selectively excite either the $SH_{1, slow}$ or $SH_{2, slow}$ mode at the same frequency as conventional single layer resonators, while suppressing the corresponding fast shear modes, without degrading $k^2_{eff}$, as shown in **Fig. 1e**. Reversing the crystallographic orientation of the upper or lower layer changes the sign of the relevant piezoelectric coefficients while preserving the elastic properties. This enables constructive electromechanical coupling across the stack and selective excitation of the higher-order shear $SH_{2, slow}$ mode, while the in-band fast shear modes are suppressed.

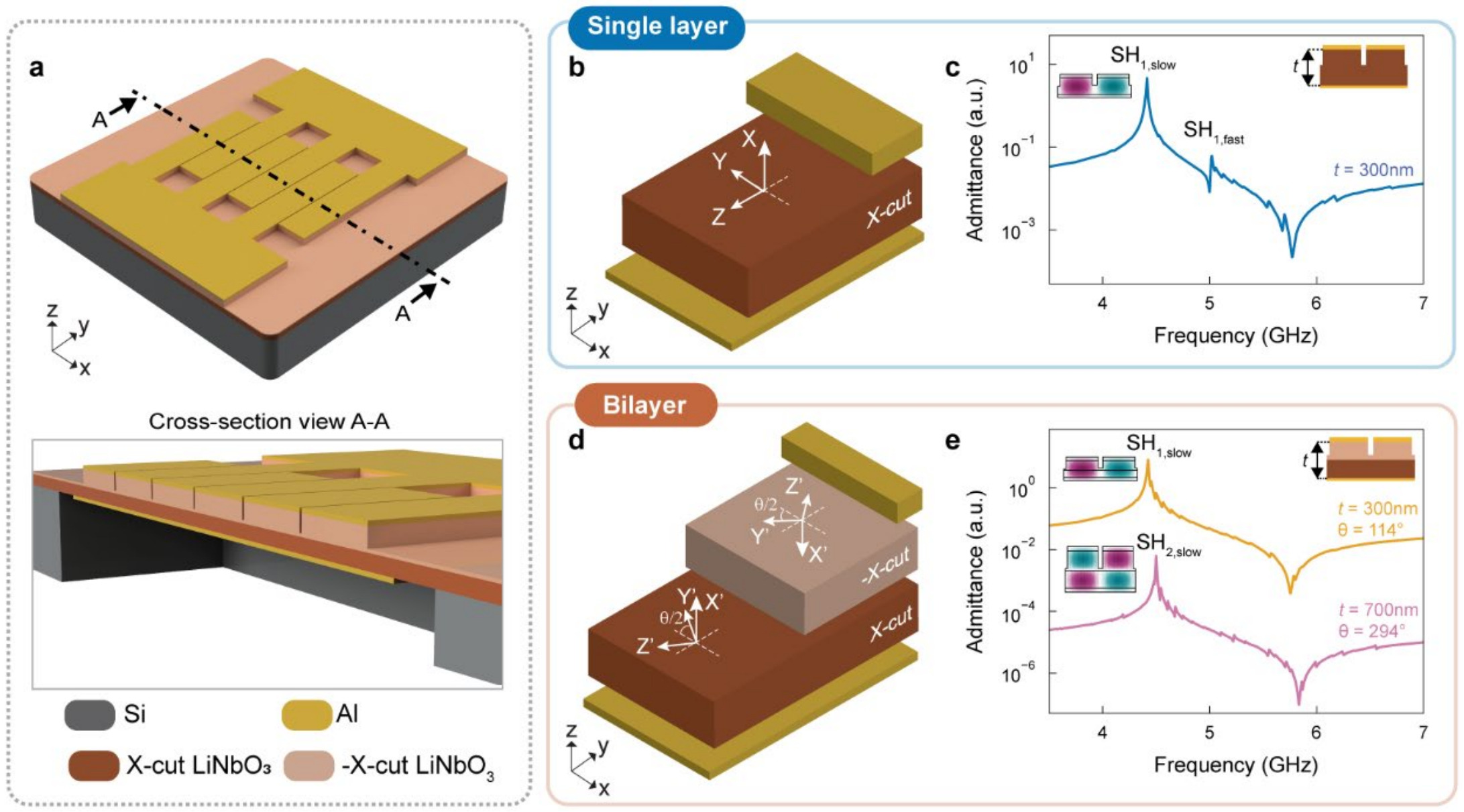

**Fig. 1 Thickness-shear bulk acoustic resonator architectures and representative responses. a** Three-dimensional schematic of the proposed device, consisting of aluminium IDTs patterned on a suspended $LiNbO_3$ membrane. The cross-section along A-A highlights the suspended structure and the floating bottom electrode enabling thickness-shear excitation. **b** Single layer X-cut $LiNbO_3$ resonator stack. **c** Admittance response of the X-cut $LiNbO_3$ thickness-shear BAW resonator, with strong coupling of the $SH_{1, slow}$ mode accompanied by the spurious $SH_{1, fast}$ mode. **d** Proposed bilayer resonator stack consisting of bonded X-cut and -X-cut $LiNbO_3$ layers with an optimized bonding angle θ. **e** Representative admittance responses demonstrating selective excitation of the $SH_{1, slow}$ and $SH_{2, slow}$ modes at approximately 4.5 GHz, together with suppression of the in-band fast shear modes. This is achieved using different $LiNbO_3$ layer thicknesses and bonding angles.

## Twisting angle for mode selection

Beyond the individual layer thicknesses, the bonding angle between the two $LiNbO_3$ layers constitutes a key design parameter. This angle must be defined prior to bonding since it modifies the piezoelectric coefficients of each layer through crystallographic rotation.

The crystallographic orientations of the X-cut and -X-cut $LiNbO_3$ layers, with in-plane rotation angles $\varphi_1$ and $\varphi_2$ respectively, are represented in **Fig. 2a**. **Fig 2b** shows the evolution of the $e_{34}$ coefficient as a function of the in-plane rotation angle $\varphi$ for X-cut and -X-cut $LiNbO_3$, together with a schematic representation of the corresponding crystallographic rotations. At $\varphi \approx 33°$, the magnitude of the $e_{34}$ coefficient is maximized, with the X-cut and -X-cut orientations exhibiting coefficients of equal magnitude and opposite sign. Once the two layers are bonded, their relative crystallographic orientation is described by the bonding angle $\theta$, as illustrated in **Fig. 2c**. We define $\theta$ as the directed angle between the ($Y'_{cr}$) axis of the bottom X-cut $LiNbO_3$ layer and that of the top -X-cut $LiNbO_3$ layer. With the convention shown in **Fig. 2c**, the bonding angle is related to the individual layer rotations by $\theta = 360 - (\varphi_1 + \varphi_2)$.

Four representative bonding configurations are shown in **Fig. 2d** with their corresponding simulated admittance response. **(i)** For $\theta = 0°$, a non-optimal configuration is obtained with both modes being excited with reduced coupling, as the piezoelectric coefficients are neither equal nor opposite. **(ii)** At $\theta = 114°$, only the $SH_{1,\ slow}$ mode is excited, this is the configuration where the $e_{34}$ and $e_{35}$ coefficients in the top layer and bottom layer are equal in both sign and magnitude. **(iii)** The optimal configuration is obtained for $\theta = 294°$ (equivalent to -66°), where $e_{34}$ coefficients have equal magnitude and opposite sign, maximizing constructive interference for the $SH_{2,\ slow}$ mode while suppressing the $SH_{1,\ slow}$ mode. The displacement and shear stress distributions at resonance confirm the excitation of the $SH_{2,\ slow}$ mode. A spurious free $SH_{2,\ slow}$ resonance at 5.9 GHz with $k^2_{eff}$ reaching 37% is achieved, highlighting the ability of the proposed configuration to combine large coupling with in-band spurious suppression at high frequency. **(iv)** In contrast, the case where $\theta = 66°$ illustrates the importance of the bonding direction: despite having the same absolute angular separation as the $\theta = 294°$ configuration, both modes are again excited with suboptimal $k^2_{eff}$.

The origin of this mode selectivity can be understood from the displacement and shear stress distributions of the $SH_{1,\ slow}$ and $SH_{2,\ slow}$ modes across the material stack under the metallized areas, shown in **Fig. 2e.** The electromechanical coupling is governed by the overlap between the stress field and the electrically induced strain field across the stack thickness. In a bilayer configuration, reversing the sign of the relevant piezoelectric coefficients enables constructive or destructive interference of the electromechanical coupling, depending on the symmetry of the stress profile. The equal and opposite $e_{34}$ coefficients obtained for the $\theta = 294°$ configuration therefore promote constructive interference for the $SH_{2,slow}$ mode (antisymmetric stress distribution) and destructive interference for the $SH_{1,slow}$ mode (symmetric stress distribution).

The evolution of this mode selectivity over the complete bonding angle range is shown in **Fig. 2f**, which presents the extracted simulated electromechanical coupling of the $SH_{1,slow}$ and $SH_{2,slow}$ modes as a function of the bonding angle $\theta$. The coupling evolutions of the two modes are phase-shifted by 180°, meaning that the maximum coupling of one mode coincides with the minimum coupling of the other. Furthermore, we notice that different maximum coupling values are reached by the two modes: the $SH_{1,slow}$ mode exhibits a maximum $k^2_{eff}$ of 43%, while the $SH_{2,slow}$ reaches 37.2%. The different maximum coupling values reached by the two modes mainly result from their different resonance frequencies and stress distributions. For comparison, a single layer X-cut $LiNbO_3$ resonator resonating in the $SH_{1,\ slow}$ around 6 GHz with identical electrode thicknesses would achieve approximately 34% coupling (see **Fig. S1b** in Supplementary Information), approximately 10% lower than the 37.2% obtained for the bilayer $SH_{2, slow}$ mode at a similar frequency.

An additional feature of the bilayer architecture is the complete suppression of the fast shear modes. Interestingly, this effect is irrespective of the bonding angle. Its persistence across different piezoelectric coefficient combinations indicates that the disappearance of the fast shear modes cannot be attributed to electromechanical coupling cancellation or to a specific choice of piezoelectric coefficients alone. Instead, this behaviour points to a modification of the excitation conditions imposed by the bilayer stack. To investigate this mechanism, the system was modelled analytically as a two-dimensional bilayer stack with traction-free boundaries and continuity of traction and displacement at the interface, where traction denotes the component of the stress tensor normal to the surface. Solving the corresponding eigenvalue problem shows that both slow and fast shear modes remain mechanically supported by the bilayer structure, as in monolayer structures. The suppression therefore originates from the excitation conditions of the modes rather than from the absence of a mechanical eigenmode. Further analysis reveals that the excitation is orthogonal to fast modes, resulting in zero coupling. A detailed formulation of the analytical model, eigenfrequency solutions and the electrical excitability criteria is provided in the Supplementary Information Section 2.

Finally, the transverse and longitudinal spurious-mode behaviour of the bilayer architecture is also investigated and discussed in Supplementary Information Section 3. The analysis reveals that the prominence of transverse and longitudinal spurious modes is complementary, preventing simultaneous suppression through in-plane rotation alone. Accordingly, the in-plane orientation of the resonators was kept constant throughout this work, while the bonding angle between the $LiNbO_3$ layers remains the focus of this work for controlling thickness-shear mode excitation.

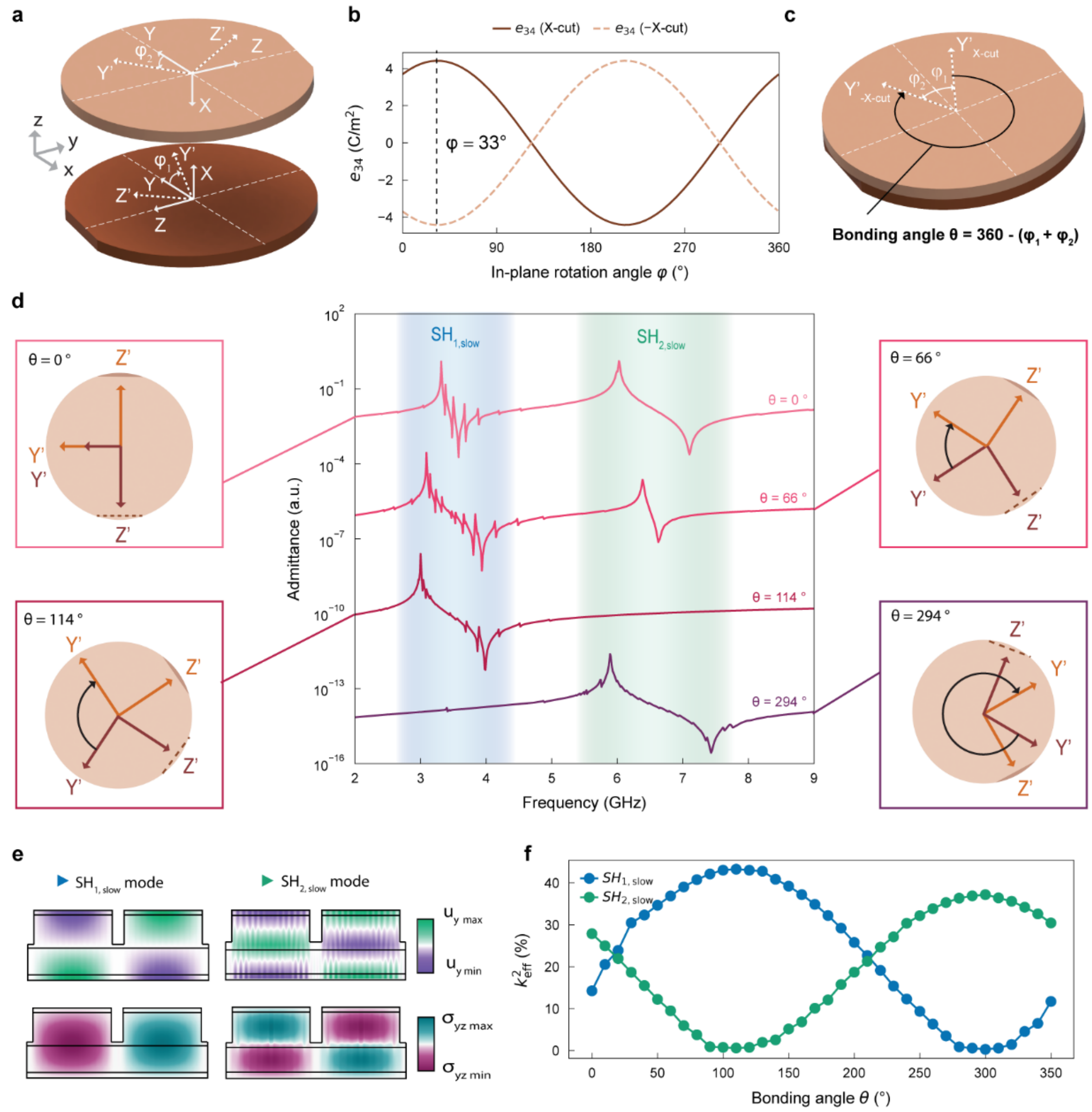

**Fig. 2 Concept and mode selective behaviour of a bilayer resonator. a** Schematic representation of the crystallographic orientations of the individual X-cut and −X-cut $LiNbO_3$ layers, defined by the in-plane rotation angles $\varphi_1$ and $\varphi_2$, respectively. **b** Evolution of the $e_{34}$ piezoelectric coefficient for X-cut and -X-cut $LiNbO_3$ as a function of the individual in-plane rotation angle φ. At $\varphi \approx 33°$, the two orientations exhibit coefficients of equal magnitude and opposite sign. ***c*** Definition of the bonding angle θ as the directed angle between the $Y'_{cr}$ axes of the bottom X-cut and top −X-cut $LiNbO_3$ layers, with $\theta = 360 - (\varphi_{1+} \varphi_2)$. **d** Simulated admittance response for selected bonding configurations: **(i)** θ = 0° corresponding to aligned Y axes, resulting in weaker excitation of both modes; **(ii)** θ = 114°, maximizing $SH_{1,\,slow}$ coupling while suppressing the $SH_{2,\,slow}$ mode **(iii)** θ = 294° (equivalent to -66°) leading to full suppression of the $SH_{1,\,slow}$ mode and maximization of $SH_{2,\,slow}$ coupling, **(iv)** θ = 66°, illustrating the importance of the twist direction to selectively exciting only the $SH_{2,\,slow}$ mode. Cases **(ii)** and **(iii)** are accompanied by one representative in-plane displacement component and one shear stress component, which are sufficient to identify the $SH_{1,\,slow}$ and $SH_{2,\,slow}$ mode shapes depending on the bonding angle. The fast shear modes are suppressed for all twisting combinations. **e** Representative in-plane displacement and shear stress components of the $SH_{1,\,slow}$ and $SH_{2,\,slow}$ modes across the resonator thickness, illustrating their respective mode symmetries. **f** Simulated electromechanical coupling factor $k^2_{eff}$ of the $SH_{1,\,slow}$ and $SH_{2,\,slow}$ modes as a function of the bonding angle θ between the two $LiNbO_3$ layers. All simulations were performed using symmetric lithium niobate layer thicknesses of 250 nm and aluminium electrode thicknesses of 75 nm.

## Device characterization

Bilayer resonators are fabricated on a bonded X-cut/-X-cut $LiNbO_3$ on Si wafer (NGK insulators) using a process adapted from previously reported YBAR technology [36]. Details of the fabrication are provided in the Materials and Methods section. **Fig. 3a** shows the optical micrograph of a completed device, where the suspended $LiNbO_3$ membrane and the backside floating electrode are indicated. To verify the actual bilayer thicknesses after trimming, the processed chip was cleaved and examined by cross-sectional SEM (see **Fig. 3b**), yielding thicknesses of 220 nm and 250 nm, for top and bottom layers respectively. The trench depth used for the devices is chosen to be 300 nm, extending into the bottom X-cut $LiNbO_3$ layer.

**Fig. 3c** shows the measured admittance response of the fabricated bilayer thickness-shear BAW resonator, along with the simulated admittance response. No de-embedding was applied to the reported measurements. The simulation incorporates the measured series resistance (2.6 Ω) to account for resistive losses. A dominant $SH_{2, slow}$ resonance is observed at 6.04 GHz in the measurement, exhibiting a strong $k^2_{eff}$ of 34.6% and no fast shear mode is present in the passband. This experimental result validates the symmetry-engineered bilayer concept by demonstrating selective excitation of the $SH_{2, slow}$ mode. The impedance ratio exceeds 2 decades, while the quality factors at resonance and antiresonance are measured to be $Q_r$ = 65 and $Q_{ar}$ = 80, respectively, with a Bode Q of approximately 100. The relatively low quality factors are consistent with acoustic losses associated with the electrodes and surface damage introduced during bilayer processing. The dominant $SH_{2, slow}$ mode is accompanied by a residual $SH_{1, slow}$ resonance at 4.16 GHz with a reduced coupling of approximately 0.9%. This residual excitation is attributed to the small difference between the $LiNbO_3$ layer thicknesses, which slightly perturbs the ideal cancellation condition. Simulated frequencies of both modes closely match the experimental values, and the predicted coupling factor values are slightly larger than the measured ones. The good agreement between measured and simulated responses further confirms the intended bonding angle of 294° in the fabricated wafer. In practice, thin-film transfer processes inevitably introduce thickness mismatches between the bonded layers. Controlled thickness asymmetry can also be intentionally exploited to tune the resonance frequencies required for ladder filter implementation. The influence of the thickness ratio $t_{LN,top}/t_{LN,bot}$ on the bilayer resonator response, achieved by controlled trimming of the -X-cut $LiNbO_3$ top layer, is experimentally assessed in **Fig. 3d**. The bottom X-cut layer thickness is 220 nm, while the top -X-cut layer thicknesses are 350 nm, 290 nm and 250 nm (confirmed by cross-sectional SEM and interferometry measurements). Both floating electrode and top IDTs are 75 nm thick. As the thickness asymmetry is reduced, the electromechanical coupling of the $SH_{1, slow}$ mode decreases from 8.13% down to 0.9%, while the $SH_{2, slow}$ coupling increases from 32.2% to 34.6%. This complementary trend is consistent with the symmetry based P3F mechanism: strong asymmetry perturbs the constructive/destructive interference condition required for selective $SH_{2, slow}$ excitation. Because the resonance frequency is primarily governed by the total piezoelectric thickness, the resonance frequency of both modes increase as the total stack thickness decreases when approaching the symmetry condition. The measured Bode Q decreases from approximately 170 to 70 as the top layer is progressively thinned.

**Fig. 3e** and **3f** show the simulated evolution of the resonance frequency and $k^2_{eff}$ of both $SH_{1, slow}$ and $SH_{2, slow}$ modes as a function of the thickness ratio $t_{LN,top}/t_{LN,bot}$. The measured performance metrics are also reported on those graphs. The resonance frequencies decrease monotonically with increasing total piezoelectric thickness. The electromechanical coupling strongly depends on the relative thicknesses of the two $LiNbO_3$ layers, with the coupling of the $SH_{1, slow}$ mode decreasing while that of the $SH_{2, slow}$ mode increases as the bilayer approaches a balanced configuration. The $SH_{2, slow}$ coupling reaches its maximum around $t_{LN,top}/t_{LN,bot}$ = 1 and remains close to its maximum over a relatively broad range of thickness ratios. The measured configurations reported in **Fig. 3e** and **3f** show good agreement between measurements and simulations. The influence of the electrode asymmetry is further discussed in the Supplementary Information Section 4.

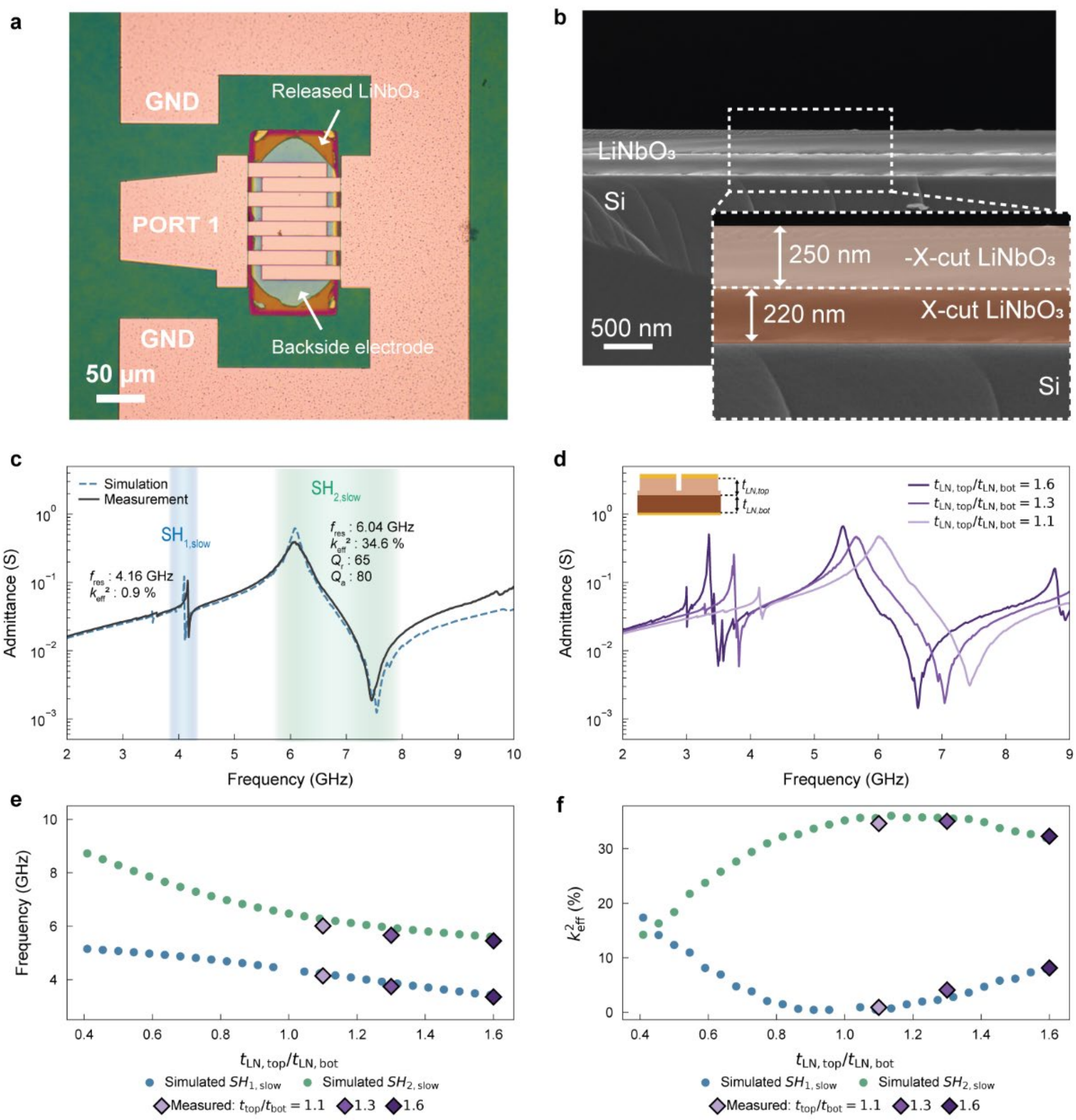

**Fig. 3 Micrographs of fabricated bilayer resonators, measured admittance response and influence of bilayer thickness symmetry. a** Optical micrograph of a fabricated bilayer device, highlighting the suspended membrane region and the backside floating electrode. **b** Cross-sectional SEM image of the bilayer stack showing individual $LiNbO_3$ thicknesses on top of the silicon substrate. **c** Measured and simulated admittance responses of the fabricated bilayer resonator. A dominant $SH_{2,\,slow}$ resonance is measured at 6.04 GHz with 34.6% electromechanical coupling. A residual $SH_{1,\,slow}$ mode is observed at 4.16 GHz with a reduced coupling of 0.9%, consistent with the asymmetry in the bilayer thicknesses. The simulation response accounts for resistive losses and is in agreement with the measured response. **d** Measured admittance responses of bilayer resonators composed of an approximately 220 nm thick X-cut $LiNbO_3$ bottom layer combined with -X-cut top layers of thicknesses 350 nm, 290 nm and 250 nm. Progressive trimming of the top layer reduces the excitation of the $SH_{1,\,slow}$ mode while enhancing the coupling of the $SH_{2,\,slow}$ mode. Simulated evolution of the resonance frequencies (**e**) and the $k^2_{eff}$ (**f**) of both modes as a function of the thickness ratio $t_{LN,top}/t_{LN,bot}$ with reported measured performances. The measured responses show good agreement with simulations.

## Filter results

To evaluate the applicability of the bilayer platform to RF filtering, proof-of-concept ladder-type filters were implemented using bilayer $LiNbO_3$ resonators. The initial design targeted operation in the 6 GHz Wi-Fi 6E band, leveraging the large coupling and spurious free response of the $SH_2$ mode to enable fractional bandwidths (FBW) exceeding 20%. The series and shunt resonators were initially designed with X-cut / −X-cut $LiNbO_3$ thickness combinations of 250/185 nm and 250/335 nm, respectively, to provide the required frequency separation. However, deviations from the nominal layer thicknesses occurred during fabrication. Since only the total bilayer thickness can be measured non-destructively, the individual layer thicknesses could only be accurately determined from cross-sectional analysis after device fabrication, which revealed a bottom X-cut layer thickness of approximately 220 nm. Based on the measured total thicknesses, the corresponding layer combinations were estimated to be approximately 220/215 nm and 220/365 nm. To partially compensate for the resulting modal imbalance, an additional 30 nm Al layer was deposited on the backside prior to device release. Although this improved modal balance, it slightly affected the coupling and increased mass loading which reduced the centre frequency and slightly decreased the electromechanical coupling.

An optical micrograph of a fabricated fifth-order ladder type filter is shown in **Fig. 4a**. The synthesized response, presented in **Fig. 4b**, is obtained using resonator parameters extracted from the cross-sectional measurements together with the measured series resistance (1.9 Ω) to account for resistive losses. The response predicts a centre frequency at 6.1 GHz with a FBW of 18.2%, demonstrating that the proposed bilayer platform can support wideband RF filtering around 6 GHz.

The measured response of the fifth order ladder-type filter is presented in **Fig. 4c,** while the third-order filter response can be found in the Supplementary Information Section 5. An experimental FBW of 18.6% centred around 5.8 GHz is obtained, in good agreement with the synthesized response. The slightly larger measured bandwidth mainly originates from the downward shift of the centre frequency, which reduces the influence of the left-hand transmission zero on the passband. The fifth order filter improves out-of-band rejection, at the expense of increased insertion loss, as expected. The measured insertion loss of 4.9 dB is primarily limited by the moderate resonator quality factors, and improvements are expected through higher-Q and larger $k^2_{\mathrm{eff}}$ resonators.

Despite the non-optimal filter performance resulting from the fabrication deviations and associated modal imbalance, these results experimentally demonstrate the feasibility of wideband RF filtering based on symmetry-engineered bilayer $LiNbO_3$ thickness-shear resonators and validate the proposed architecture at the proof-of-concept level. To the best of our knowledge, this represents the first ladder-type filter based on vertically excited P3F $LiNbO_3$ thickness-shear resonators.

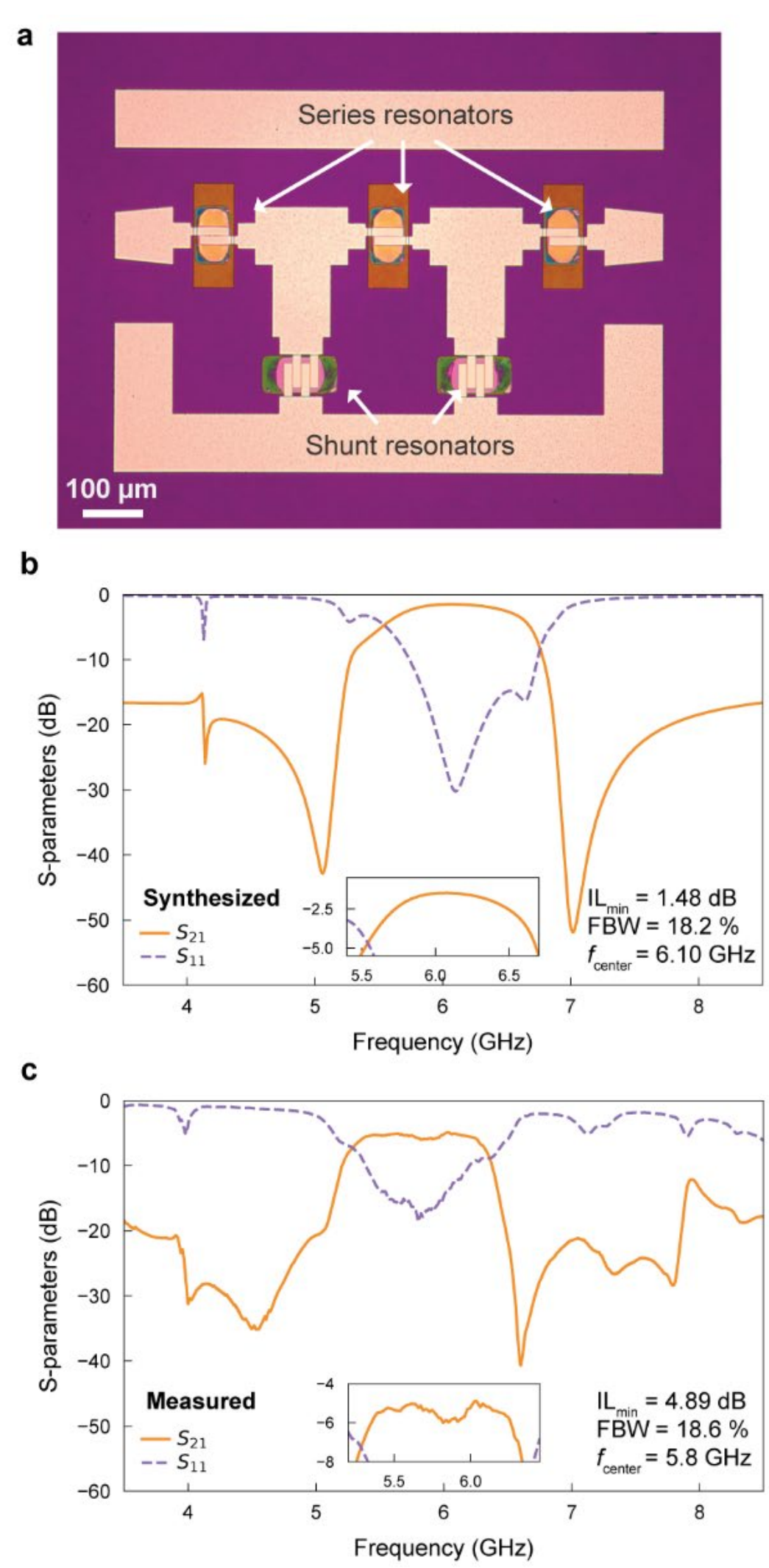

**Fig. 4 Filter implementation using bilayer resonators. a** Optical micrograph of a fabricated 5th order ladder-type filter. The top layer $LiNbO_3$ was selectively trimmed for the series resonators to achieve the required frequency separation. **b** Synthesized response of a 5th order ladder-type filter using simulated bilayer resonators with 220 nm/215 nm and 220 nm/365 nm X-cut/-X-cut $LiNbO_3$ thickness combinations, obtained by experimentally measuring the layer thicknesses. **c** Measured response of a 5th order ladder-type filter showing 18.6% of FBW centred around 5.8 GHz. No de-embedding was applied to the measured results.

# Discussion

This work demonstrates that P3F-based symmetry engineering introduces a new degree of freedom for the design of thickness-shear bulk acoustic resonators. While conventional acoustic resonator design primarily relies on the selection of the piezoelectric material, crystal orientation, layer thickness and electrode configuration, the present approach exploits the symmetry of vertically stacked piezoelectric films as an additional design parameter to control the electromechanical excitation of acoustic modes. This enables the selective enhancement or suppression of slow thickness shear acoustic modes through the bonding angle between the two piezoelectric layers, extending the design space beyond that of conventional single-layer resonators.

Conventional single layer X-cut YBAR/FBAR resonators simultaneously excite both the fundamental $SH_{1,slow}$ and the $SH_{1,fast}$ modes preventing practical filter implementation. Although Y-cut and YX 163° $LiNbO_3$ films provide a cleaner spectral response, their electromechanical coupling at 6 GHz remains limited to approximately 26% (see Fig. S1). In contrast, the proposed bilayer architecture combines the high electromechanical coupling associated with X-cut $LiNbO_3$ with selective $SH_{2, slow}$ excitation, enabling spurious-free operation at higher frequencies while retaining a comparatively thick piezoelectric stack.

Although both the $SH_{1, slow}$ and $SH_{2, slow}$ can be excited spurious-free with the bilayer architecture by choosing the appropriate bonding angle (see **Fig. 1e**), the $SH_{2, slow}$ mode offers the greatest practical interest for RF filtering applications. Exploiting the higher order mode allows operation at higher frequencies without aggressive thickness reduction, potentially improving both the power handling and the mechanical robustness. At 6 GHz, the measured coupling of approximately 35% is competitive with values reported for $A_1$ XBAR [31,34] and multilayer devices [42] at similar or lower frequencies. Compared with $SH_0$ and $S_0$ plate-wave resonators, the thickness-defined nature of the proposed resonators simplifies frequency scaling, while the use of X-cut $LiNbO_3$ enables substantially larger electromechanical coupling [23]. The proof-of-concept ladder filters demonstrate the large FBW achievable with this architecture, while also revealing the sensitivity of modal excitation to slight asymmetries in the stack.

Compared to earlier demonstrations of similar concepts in the MHz regime, scaling to 6 GHz imposes tighter fabrication tolerances, illustrating both the potential and the challenges of nanoscale symmetry engineering. The residual $SH_{1, slow}$ response originates from nanoscale thickness mismatches between the two bonded films, highlighting the sensitivity of this architecture to vertical symmetry. Moreover, the observed reduction in quality factor with increasing trimming depth suggests increased acoustic losses, likely arising from enhanced surface damage or stronger interaction of the acoustic field with the electrodes and surface regions. The ladder filter implementation further illustrates the trade-off between achieving the required frequency separation and maintaining optimal modal purity. Controlled thickness asymmetry is required to introduce the frequency offset between the series and shunt resonators, but it simultaneously perturbs the ideal symmetry conditions for selective $SH_{2, slow}$ excitation.

Future work will focus on improved layer thickness control and electrode optimization to further improve Q factor and fully suppress residual $SH_{1, slow}$ mode. The thicker piezoelectric stack additionally provides opportunities to push operation toward higher frequencies while maintaining strong coupling, positioning bilayer X-cut resonators as a promising platform for next-generation wideband RF filtering. More broadly, these results establish vertical symmetry engineering as a powerful mechanism for controlling electromechanical coupling excitation in thickness-shear resonators. Beyond the bilayer X-cut $LiNbO_3$ architecture demonstrated here, this concept could be extended to other $LiNbO_3$ cuts and higher-order acoustic modes, opening new opportunities for wideband microwave acoustic filters.

# Materials and Methods

## Materials

The devices were fabricated on a bonded lithium niobate bilayer wafer provided by NGK Insulators, consisting of a 250 nm X-cut $LiNbO_3$ layer and a 350 nm -66° -X-cut $LiNbO_3$ layer, bonded onto a high-resistive silicon wafer. The total piezoelectric thickness varied between 570 nm and 630 nm across the wafer. Platinum alignment marks were first defined on the front side of the wafer. After thinning the silicon substrate to 250 µm, backside Pt alignment marks were deposited on the backside surface before dicing the wafer into individual chips.

## Device fabrication

The process was adapted from the previously reported single-layer YBAR process [35], with additional steps enabling local thinning of the top $LiNbO_3$ layer and trench etching within the bilayer stack. An asymmetric lithium niobate stack was intentionally selected to enable resonance frequency tuning, as predicted by finite-element simulations.

*Frontside processing*

The top $LiNbO_3$ surface was locally thinned by low power ion beam etching (IBE, Veeco Nexus IBE350) to minimize surface damage. A Cr/Al (5/75 nm) metal stack was then deposited by electron-beam evaporation (Bühler Alzenau LTE 700). Resonator electrodes are defined by electron-beam lithography (Raith EBPG5000+) using ZEP520A e-beam resist, followed by ion beam etching of both the metal and the $LiNbO_3$ layers. The trenches were etched to a depth of 300 nm. Following resist removal by $O_2$ plasma and heated Remover 1165, 500 nm-thick aluminium contact pads were defined by lift-off using direct laser writing (Heidelberg MLA 150).

*Backside processing*

The frontside of the chip was protected with a 5 µm thick photoresist layer before temporary bonding to a carrier wafer using QS135 wax. Backside photolithography was carried out using a 6 µm thick positive tone photoresist layer (Merck AZ10XT), patterned by direct laser writing with alignment to the frontside features ensured by infrared microscopy. The silicon substrate was subsequently etched by deep reactive ion etching (DRIE, Alcatel AMS 200 SE) using a Bosch process to release the suspended membrane. A floating backside electrode consisting of Cr/Al (5/75 nm) was then deposited by electron-beam evaporation.

*Device release*

The protective photoresists were removed in heated Remover 1165, followed by rinsing in IPA and air drying to release the suspended bilayer resonators. An optical micrograph of a completed device is shown in **Fig. 3a**.

## Electrical characterization

The electrical performance of the fabricated bilayer devices was measured in air using a one-port GSG configuration and measured with an R&S®ZNB20 vector network analyser (VNA) with an input power of -15 dBm. No electrical de-embedding was applied.

The measured admittance responses were fitted using a modified Butterworth–Van Dyke (mBVD) equivalent circuit model to extract the resonator parameters.

For filter characterization, the devices were measured in a two-port configuration using GSG probes and the transmission response ($S_{21}$) and reflection ($S_{11}$) were measured using the same VNA.

The definition used for the effective electromechanical coupling coefficient $k_{\mathrm{eff}}^2$ is:

$$k_{eff}^2 = \frac{f_{ar}^2 - f_r^2}{f_{ar}^2}$$

where $f_r$ and $f_{ar}$ correspond to the resonance and antiresonance frequencies, respectively [47].

### Acknowledgements

The authors gratefully acknowledge the Center of MicroNanoTechnology (CMi) at EPFL for its support and assistance in the fabrication of the devices as well as NGK Insulators for providing the bilayer $LiNbO_3$ thin-film-on-Si wafer.

### Author contributions

F.H. designed and investigated the resonators and filters, performed the simulations, fabrication, measurements and data analysis, and wrote the manuscript. S.S. contributed to the theoretical analysis and simulations and reviewed the manuscript. L.G.V. supervised the work, acquired funding and reviewed the manuscript.

### Data availability

The data supporting the findings of this study are available from the corresponding author upon reasonable request.

### Conflict of Interest

The authors declare no competing interests.

## References

1. Andrews, J. G. *et al.* What Will 5G Be? *IEEE J. Sel. Areas Commun.* **32**, 1065–1082 (2014).
2. Gupta, S., Mehdizadeh, E., Cheema, K. & Shealy, J. B. Miniaturized Ultrawide Bandwidth WiFi 6E Diplexer Implementation Using XBAW RF Filter Technology. in *2022 IEEE/MTT-S International Microwave Symposium - IMS 2022* 880–882 (IEEE, Denver, CO, USA, 2022). doi:10.1109/IMS37962.2022.9865319.
3. Ruby, R. A Snapshot in Time: The Future in Filters for Cell Phones. *IEEE Microw. Mag.* **16**, 46–59 (2015).
4. Takai, T. *et al.* High-Performance SAW Resonator With Simplified $LiTaO_3$ /$SiO_2$ Double Layer Structure on Si Substrate. *IEEE Trans. Ultrason. Ferroelectr. Freq. Control* **66**, 1006–1013 (2019).
5. Luo, T. *et al.* Shear horizontal mode surface acoustic wave resonator with ultra-high electromechanical coupling based on the X-cut LNOI acoustic platform. *Appl. Phys. Lett.* **126**, 192204 (2025).
6. Tanaka, S., Guo, Y. & Kadota, M. Evolution of SAW and BAW Devices Using Thin $LiTaO_3$ and $LiNbO_3$. in *2024 IEEE MTT-S International Conference on Microwave Acoustics & Mechanics (IC-MAM)* 169–172 (IEEE, Chengdu, China, 2024). doi:10.1109/IC-MAM60575.2024.10538491.
7. Su, R. *et al.* Over GHz bandwidth SAW filter based on 32° Y-X LN/$SiO_2$/poly-Si/Si heterostructure with multilayer electrode modulation. *Appl. Phys. Lett.* **120**, 253501 (2022).
8. Aigner, R. *et al.* BAW Filters for 5G Bands. in *2018 IEEE International Electron Devices Meeting (IEDM)* 14.5.1-14.5.4 (IEEE, San Francisco, CA, 2018). doi:10.1109/IEDM.2018.8614564.
9. Liffredo, M., Xu, N., Stettler, S., Peretti, F. & Villanueva, L. G. Piezoelectric and elastic properties of $Al_{0.60}Sc_{0.40}N$ thin films deposited on patterned metal electrodes. *J. Vac. Sci. Technol. A* **42**, 043404 (2024).

10. Moe, C. *et al.* Highly Doped AlScN 3.5 GHz XBAW Resonators with 16% $k^2_{eff}$ for 5G RF Filter Applications. in *2020 IEEE International Ultrasonics Symposium (IUS)* 1–4 (IEEE, Las Vegas, NV, USA, 2020). doi:10.1109/IUS46767.2020.9251412.
11. Zou, Y. *et al.* Aluminum scandium nitride thin-film bulk acoustic resonators for 5G wideband applications. *Microsyst. Nanoeng.* **8**, 124 (2022).
12. Schaffer, Z. & Piazza, G. Investigation of Damping and Ladder Filter Synthesis for 3 GHz 20% Scandium-Doped Aluminum Nitride Cross-Sectional Lame Mode Resonators. in *2020 IEEE International Ultrasonics Symposium (IUS)* 1–4 (IEEE, Las Vegas, NV, USA, 2020). doi:10.1109/IUS46767.2020.9251790.
13. Zhou, C. *et al.* Highly Doped Single Crystal $Al_{1-x}Sc_xN$ Bulk Acoustic Resonators for High-Frequency and Wideband Applications. *IEEE Trans. Electron Devices* **71**, 6329–6335 (2024).
14. Zhao, W. *et al.* X-band epi-BAW resonators. *J. Appl. Phys.* **132**, 024503 (2022).
15. Izhar *et al.* Periodically poled aluminum scandium nitride bulk acoustic wave resonators and filters for communications in the 6G era. *Microsyst. Nanoeng.* **11**, 19 (2025).
16. Vetury, R. *et al.* A Manufacturable AlScN Periodically Polarized Piezoelectric Film Bulk Acoustic Wave Resonator (AlScN P3F BAW) Operating in Overtone Mode at X and Ku Band. in *2023 IEEE/MTT-S International Microwave Symposium - IMS 2023* 891–894 (IEEE, San Diego, CA, USA, 2023). doi:10.1109/IMS37964.2023.10188141.
17. Kochhar, A. *et al.* X-band Bulk Acoustic Wave Resonator (XBAW) using Periodically Polarized Piezoelectric Films (P3F). in *2023 IEEE International Ultrasonics Symposium (IUS)* 1–4 (IEEE, Montreal, QC, Canada, 2023). doi:10.1109/IUS51837.2023.10306825.
18. Xu, H. *et al.* SAW Filters on $LiNbO_3$ /SiC Heterostructure for 5G n77 and n78 Band Applications. *IEEE Trans. Ultrason. Ferroelectr. Freq. Control* **70**, 1157–1169 (2023).
19. Zheng, P. *et al.* Near 5-GHz Longitudinal Leaky Surface Acoustic Wave Devices on $LiNbO_3$/SiC Substrates. *IEEE Trans. Microw. Theory Tech.* 1–9 (2023) doi:10.1109/TMTT.2023.3305078.
20. Zhou, H. *et al.* Ultrawide-Band SAW Devices Using $SH_0$ Mode Wave with Increased Velocity for 5G Front-Ends. in *2021 IEEE International Ultrasonics Symposium (IUS)* 1–4 (IEEE, Xi'an, China, 2021). doi:10.1109/IUS52206.2021.9593689.
21. Chulukhadze, V. *et al.* 2 to 16 GHz Fundamental Symmetric Mode Acoustic Resonators in Piezoelectric Thin-film Lithium niobate.
22. Pop, F. V., Kochhar, A. S., Vidal-Alvarez, G. & Piazza, G. Laterally vibrating lithium niobate MEMS resonators with 30% electromechanical coupling coefficient. in *2017 IEEE 30th International Conference on Micro Electro Mechanical Systems (MEMS)* 966–969 (IEEE, Las Vegas, NV, USA, 2017). doi:10.1109/MEMSYS.2017.7863571.
23. Tetro, R., Colombo, L., Gubinelli, W., Giribaldi, G. & Rinaldi, M. X-Cut Lithium Niobate $S_0$ Mode Resonators for 5G Applications. in *2024 IEEE 37th International Conference on Micro Electro Mechanical Systems (MEMS)* 1102–1105 (IEEE, Austin, TX, USA, 2024). doi:10.1109/MEMS58180.2024.10439491.
24. Faizan, M. & Villanueva, L. G. Frequency-scalable fabrication process flow for lithium niobate based Lamb wave resonators. *J. Micromechanics Microengineering* **30**, 015008 (2020).
25. Kadota, M., Esashi, M., Tanaka, S., Kuratani, Y. & Kimura, T. High frequency resonators with wide bandwidth using $SH_0$ mode plate wave in thin $LiNbO_3$. in *2013 IEEE International Ultrasonics Symposium (IUS)* 1680–1683 (IEEE, Prague, Czech Republic, 2013). doi:10.1109/ULTSYM.2013.0428.
26. Stettler, S. & Villanueva, L. G. Suspended lithium niobate acoustic resonators with Damascene electrodes for radiofrequency filtering. *Microsyst. Nanoeng.* **11**, 131 (2025).

27. Wu, S. *et al.* Large Coupling and Spurious-Free $SH_0$ Plate Acoustic Wave Resonators Using $LiNbO_3$ Thin Film. *IEEE Trans. Electron Devices* 1–8 (2023) doi:10.1109/TED.2023.3297561.
28. Zou, J. *et al.* Ultra-Large-Coupling and Spurious-Free SH0 Plate Acoustic Wave Resonators Based on Thin LiNbO3. *IEEE Trans. Ultrason. Ferroelectr. Freq. Control* **67**, 374–386 (2020).
29. Stettler, S., Navarro-Gessé, E., Collado, C., Mateu, J. & Villanueva, L. G. A bulk acoustic resonator with vertical electrodes for wideband filters. *Commun. Eng.* https://doi.org/10.1038/s44172-026-00733-1 (2026) doi:10.1038/s44172-026-00733-1.
30. Lu, R., Yang, Y., Link, S. & Gong, S. A1 Resonators in 128° Y-cut Lithium Niobate with Electromechanical Coupling of 46.4%. *J. Microelectromechanical Syst.* **29**, 313–319 (2020).
31. Tong, X. *et al.* High Figure-of-Merit $A_1$-Mode Lamb Wave Resonators Operating Around 6 GHz Based on the $LiNbO_3$ Thin Film. *J. Phys. Appl. Phys.* https://doi.org/10.1088/1361-6463/ad3e06 (2024) doi:10.1088/1361-6463/ad3e06.
32. Plessky, V. *et al.* 5 GHz laterally-excited bulk-wave resonators (XBARs) based on thin platelets of lithium niobate. *Electron. Lett.* **55**, 98–100 (2019).
33. Xie, Z. *et al.* Near-Spurious-Free 6.5 GHz XBARs With Dimension-Matched and Response-Averaged Electrodes. *J. Microelectromechanical Syst.* 1–7 (2025) doi:10.1109/JMEMS.2025.3577619.
34. Yandrapalli, S., Eroglu, S. E. K., Plessky, V., Atakan, H. B. & Villanueva, L. G. Study of Thin Film $LiNbO_3$ Laterally Excited Bulk Acoustic Resonators. *J. Microelectromechanical Syst.* **31**, 217–225 (2022).
35. Bousquet, M. *et al.* $LiNbO_3$ Film Bulk Acoustic Resonator for n79 band. in *2022 IEEE International Ultrasonics Symposium (IUS)* 1–5 (IEEE, Venice, Italy, 2022). doi:10.1109/IUS54386.2022.9957601.
36. Yandrapalli, S. *et al.* Toward Band n78 Shear Bulk Acoustic Resonators Using Crystalline Y-Cut Lithium Niobate Films With Spurious Suppression. *J. Microelectromechanical Syst.* **32**, 327–334 (2023).
37. Plessky, V. P., Koskela, J. & Yandrapalli, S. Crystalline Y-cut Lithium Niobate Layers for the Bulk Acoustic Wave Resonator (YBAR). in *2020 IEEE International Ultrasonics Symposium (IUS)* 1–4 (IEEE, Las Vegas, NV, USA, 2020). doi:10.1109/IUS46767.2020.9251647.
38. Bousquet, M. *et al.* Lithium niobate film bulk acoustic wave resonator for sub-6 GHz filters. in *2020 IEEE International Ultrasonics Symposium (IUS)* 1–4 (IEEE, Las Vegas, NV, USA, 2020). doi:10.1109/IUS46767.2020.9251654.
39. Qin, Z.-H. *et al.* Solidly Mounted Longitudinally Excited Shear Wave Resonator (YBAR) Based on Lithium Niobate Thin-Film. *Micromachines* **12**, 1039 (2021).
40. Gorisse, M. *et al.* High Frequency LiNbO $_3$ Bulk Wave Resonator. in *2019 Joint Conference of the IEEE International Frequency Control Symposium and European Frequency and Time Forum (EFTF/IFC)* 1–2 (IEEE, Orlando, FL, USA, 2019). doi:10.1109/FCS.2019.8856017.
41. Hartmann, F. *et al.* A 5G n77 Filter Using Shear Bulk Mode Resonator With Crystalline X-cut Lithium Niobate Films. in *2024 IEEE International Microwave Filter Workshop (IMFW)* 78–80 (IEEE, Cocoa Beach, FL, USA, 2024). doi:10.1109/IMFW59690.2024.10477114.
42. Kramer, J. *et al.* Trilayer Periodically Poled Piezoelectric Film Lithium Niobate Resonator. in *2023 IEEE International Ultrasonics Symposium (IUS)* 1–4 (IEEE, Montreal, QC, Canada, 2023). doi:10.1109/IUS51837.2023.10306831.
43. Lu, R. & Gong, S. A 15.8 GHz A6 Mode Resonator with Q of 720 in Complementarily Oriented Piezoelectric Lithium Niobate Thin Films. in *2021 Joint Conference of the European Frequency and Time Forum and IEEE International Frequency Control Symposium (EFTF/IFCS)* 1–4 (IEEE, Gainesville, FL, USA, 2021). doi:10.1109/EFTF/IFCS52194.2021.9604327.

44. Lu, R., Yang, Y., Link, S. & Gong, S. Enabling Higher Order Lamb Wave Acoustic Devices With Complementarily Oriented Piezoelectric Thin Films. *J. Microelectromechanical Syst.* **29**, 1332–1346 (2020).
45. Qin, Z.-H. *et al.* Double-layer LiNbO3 longitudinally excited shear wave resonators with ultra-large electromechanical coupling coefficient and spurious-free performance. *Appl. Phys. Lett.* **126**, 193503 (2025).
46. Yao, H. *et al.* Twist piezoelectricity: giant electromechanical coupling in magic-angle twisted bilayer LiNbO3. *Nat. Commun.* **15**, 5002 (2024).
47. *IEEE Standard on Piezoelectricity*. https://ieeexplore.ieee.org/document/26560/ doi:10.1109/IEEESTD.1988.79638.

# Supplementary Information: Bilayer Lithium Niobate Acoustic Resonators for Spurious-Free Wideband Operation at 6 GHz

Florian Hartmann[1], Silvan Stettler[1], Luis Guillermo Villanueva[1]

[1] Advanced Nanoelectromechanical Systems Laboratory, École Polytechnique Fédérale de Lausanne (EPFL), 1015 Lausanne, Switzerland

## Section 1 - Single layer X-cut, Y-cut & YX163° $LiNbO_3$ thickness-shear FBARs

**Fig. S1** presents the elastic and piezo-stiffened slowness curves of X-cut, Y-cut and YX163° $LiNbO_3$ together with simulated admittance responses of single layer thickness shear X-cut, Y-cut and YX163° $LiNbO_3$ FBAR resonators.

The slowness curves shown in **Fig. S1a**, **S1c** and **S1e** correspond to X-cut, Y-cut and YX163° $LiNbO_3$, respectively. Solving the Christoffel equation yields three acoustic wave solutions for both the elastic and piezo-stiffened cases, corresponding to two shear waves and one longitudinal wave. In X-cut $LiNbO_3$, both shear waves exhibit non-zero electromechanical coupling. The slow shear mode ($SH_{1,\ slow}$) exhibits strong coupling, whereas the fast shear mode ($SH_{1,\ fast}$) presents a weaker but non-negligible coupling as evidenced by the separation between the elastic and piezo-stiffened slowness curves showed in the inset. The longitudinal thickness mode exhibits no piezoelectric coupling. For Y-cut and YX163° $LiNbO_3$, only the $SH_{1,\ slow}$ is coupled, although with a lower coupling strength than in the X-cut case, as indicated by the smaller separation between the elastic and piezo-stiffened slowness curves. The $SH_{1,fast}$ curves overlap, indicating zero piezoelectric coupling. Conversely, the longitudinal mode exhibits a finite separation and therefore becomes electromechanically coupled at higher frequencies for Y-cut $LiNbO_3$.

**Fig. S1b**, **S1d** and **S1f** show the simulated admittances responses of resonators designed to operate around 4.5 GHz and 6 GHz, using $LiNbO_3$ thicknesses of 300 and 180 nm, respectively, with 75 nm thick Al electrodes. For the X-cut resonators, reducing the piezoelectric thickness shifts the resonance frequency from 4.42 GHz to 6.03 GHz while the $k^2_{\mathrm{eff}}$ drops from 41.5% to 34.5%. In addition, we observe the presence of the in-band $SH_{1,\ fast}$ mode, which is intrinsic to this crystal orientation. In contrast, the Y-cut $LiNbO_3$ resonators exhibit a spurious-free response but with lower coupling, decreasing to 26.1% at 6.17 GHz for the thinner resonator. Similarly, for YX163° $LiNbO_3$ the admittance response is clean but $k^2_{\mathrm{eff}}$ decreases from 32.3% to 26.7%. In both crystal cuts, the 180 nm thick resonators employ a 75 nm Al electrode corresponding to approximately 45% of the piezoelectric thickness. Scaling to even higher frequencies would require an even larger metal-to-piezoelectric thickness ratio, resulting in excessive mass loading and a significant degradation of the admittance response.

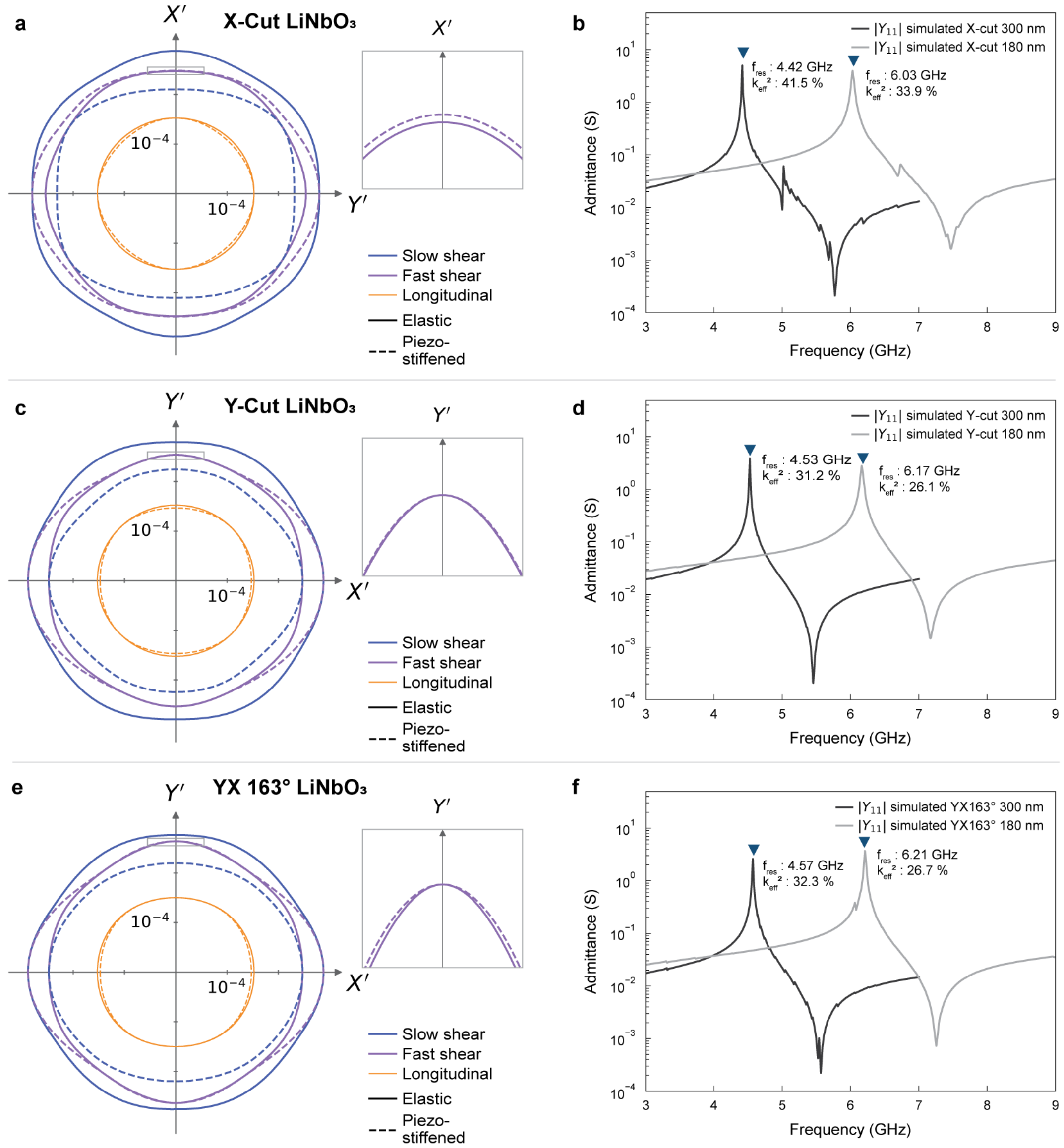

**Fig. S1 Simulated slowness curves and admittance responses of single-layer X-cut, Y-cut and YX163° $LiNbO_3$ thickness-shear FBAR resonators operating around 4.5 and 6 GHz**. **a** Elastic and piezo-stiffened slowness curves of X-cut $LiNbO_3$.The inset highlights the non-zero coupling of the $SH_{1,\ fast}$ mode. **b** Simulated admittance responses of 300 nm and 180 nm thick single layer X-cut $LiNbO_3$ thickness-shear FBAR resonators with 75 nm thick Al electrodes. The resonances at 4.42 and 6.03 GHz exhibit an $SH_{1,\ fast}$ mode, while the coupling decreases from 41.5% to 34.5% as the piezoelectric thickness is reduced. **c** Elastic and piezo-stiffened slowness curves of Y-cut $LiNbO_3$. The inset highlights the zero coupling of the $SH_{1,\ fast}$ mode. **d** Simulated admittance responses of 300 nm and 180 nm thick single layer Y-cut $LiNbO_3$ thickness-shear FBAR resonators with 75 nm thick Al electrodes. The resonances at 4.53 and 6.17 GHz are spurious free but the coupling decreases to 26.1% for the thinner piezoelectric layer case. **e** Elastic and piezo-stiffened slowness curves of YX163° $LiNbO_3$. The inset highlights the zero coupling of the $SH_{1,\ fast}$ mode. **f** Simulated admittance responses of 300 nm and 180 nm thick single layer YX163° $LiNbO_3$ thickness-shear FBAR resonators with 75 nm thick Al electrodes. The resonances at 4.57 and 6.21 GHz are spurious free but the coupling decreases to 26.7% for the thinner piezoelectric layer case.

# Section 2 - SVD explanation

To understand the physical origin of the in-band spurious mode suppression in the bilayer X-cut $LiNbO_3$ thickness-shear FBAR architecture, a one-dimensional analytical model based on acoustic wave equations was developed.

**Single layer analysis**

To begin with, the full analysis is explained at a single layer level before applying it to the bilayer architecture. We start from the linear piezoelectric constitutive equations in stress-charge form:

$$\boldsymbol{T} = c^E \boldsymbol{S} - e^T \boldsymbol{E} \qquad (1)$$
$$\boldsymbol{D} = e\boldsymbol{S} + \varepsilon^S \boldsymbol{E} \qquad (2)$$

Here we consider only the 1D problem, therefore only the $T_4$ and $T_5$ shear components and the electric displacement $D_3$ are relevant:

$$T_4 = c_{44}{}^E S_4 + c_{45}{}^E S_5 - e_{34} E_3 \qquad (3)$$

$$T_5 = c_{54}{}^E S_4 + c_{55}{}^E S_5 - e_{35} E_3 \qquad (4)$$

$$D_3 = e_{34} S_4 + e_{35} S_5 + \varepsilon_{33}{}^S E_3 \qquad (5)$$

We solve the problem under open-circuit electrical boundary conditions, therefore we can eliminate the electric field under D = 0 which ultimately leads to the stiffened elastic constants:

$$\begin{cases} c_{44}{}^* = c_{44} + \dfrac{e_{34}{}^2}{\varepsilon_{33}} \\ c_{55}{}^* = c_{55} + \dfrac{e_{35}{}^2}{\varepsilon_{33}} \\ c_{45}{}^* = c_{45} + \dfrac{e_{34} e_{35}}{\varepsilon_{33}} \\ c_{54}{}^* = c_{54} + \dfrac{e_{34} e_{35}}{\varepsilon_{33}} \end{cases} \qquad (6)$$

Using the properties of Z-cut LiNbO3 and rotating them using Euler angles in ZXZ convention, the properties of X-cut $LiNbO_3$ are obtained. Using Christoffel's equation, we can compute the slowness of the 3 acoustic wave solutions (2 shear, 1 compressional) in an infinite bulk material with stiffness, piezoelectric and permittivity tensors, with a vertical propagation direction. Solving Christoffel's equation yields two shear-wave solutions with polarization vectors $p_1$ and $p_2$, whose displacements lie in the plane perpendicular to the propagation direction (the thickness direction $z$).

For a single layer case, we can define a slab of thickness h with mechanically free surfaces at z=-h/2 and z=h/2 as represented in **Fig. S2a.** This implies that the traction vector must vanish at both surfaces:

$$\boldsymbol{T}\,\boldsymbol{n} = \boldsymbol{0} \qquad (7)$$

with **n** the outward normal. Using the non-zero shear stress components and (7), the boundary conditions become:

$$T_4\left(-\frac{h}{2}\right) = 0, \qquad T_5\left(-\frac{h}{2}\right) = 0$$
$$T_4\left(\frac{h}{2}\right) = 0, \qquad T_5\left(\frac{h}{2}\right) = 0 \qquad (8)$$

The shear displacement field is expressed as a superposition of forward and backward propagating shear waves:

$$\boldsymbol{u}(z) = \boldsymbol{p_1}\left(A_1 e^{-ik_1 z} + B_1 e^{+ik_1 z}\right) + \boldsymbol{p_2}\left(A_2 e^{-ik_2 z} + B_2 e^{+ik_2 z}\right) \quad (9)$$

with $\mathbf{p_1}$ and $\mathbf{p_2}$ the polarization vectors and $k_1$ and $k_2$ their corresponding wavenumbers. The shear strains can be defined in the following way:

$$\begin{cases} S_4(z) = \dfrac{\partial u_y}{\partial z} = ik_1 p_{1,y}\left(A_1 e^{-ik_1 z} + B_1 e^{+ik_1 z}\right) + ik_2 p_{2,y}\left(A_2 e^{-ik_2 z} + B_2 e^{+ik_2 z}\right) \\ S_5(z) = \dfrac{\partial u_x}{\partial z} = ik_1 p_{1,x}\left(A_1 e^{-ik_1 z} + B_1 e^{+ik_1 z}\right) + ik_2 p_{2,x}\left(A_2 e^{-ik_2 z} + B_2 e^{+ik_2 z}\right) \end{cases} \quad (10)$$

And finally, we can obtain $T_4$ and $T_5$ using (3), (4), (6) and (10) and substituting $E_3$ with $E_0$:

$$\begin{cases} T_4(z) = c_{44}{}^{*} S_4 + c_{45}{}^{*} S_5 - e_{34} E_0 \\ T_5(z) = c_{54}{}^{E} S_4 + c_{55}{}^{E} S_5 - e_{35} E_0 \end{cases} \quad (11)$$

$T_4$ and $T_5$ are linear combinations of the four amplitudes $A_1$,$B_1$,$A_2$ and $B_2$ and we can simplify them the following way:

$$\begin{cases} T_4{}^{mech}(z) - e_{34} E_0 = 0 \\ T_5{}^{mech}(z) - e_{35} E_0 = 0 \end{cases} \quad (12)$$

Collecting the four boundary conditions (8) we obtain a system:

$$\boldsymbol{A}\,\boldsymbol{a} = \boldsymbol{b} \quad (13)$$

with

$$A = \begin{pmatrix} T_4{}^{mech}(-h/2) \\ T_5{}^{mech}(-h/2) \\ T_4{}^{mech}(h/2) \\ T_5{}^{mech}(h/2) \end{pmatrix}, \quad a = \begin{pmatrix} A_1 \\ \mathrm{B}_1 \\ A_2 \\ B_2 \end{pmatrix}, \quad b = \begin{pmatrix} e_{34} E_0 \\ e_{35} E_0 \\ e_{34} E_0 \\ e_{35} E_0 \end{pmatrix} \quad (14)$$

Solving this system yields the 4 amplitudes that determine displacement, strain and stress fields inside the piezoelectric slab for a given frequency.

Using (5) and Gauss' law:

$$\nabla \cdot \boldsymbol{D} = 0 \ \ becomes \ \frac{dD_3}{dz} = 0 \quad (15)$$

So $D_3$ is constant along the thickness and under open-circuit conditions we can write

$$D_3(z) = D_3 = E_0 \varepsilon_{33}{}^{S} \quad (16)$$

$$E_3(z) = E_0 - \frac{e_{34} S_4(z) + e_{35} S_5(z)}{\varepsilon_{33}{}^{S}} \quad (17)$$

The voltage between the electrodes is obtained by integrating the electric field across the thickness:

$$V = -\int^{z} E_3(z)dz \quad (18)$$

We write the electrode current under harmonic excitation:

$$I(\omega) = j\omega A D_3 = j\omega A E_0 {\varepsilon_{33}}^S \qquad (19)$$

with A the area. Finally we can get the electrical impedance and normalize $E_0$ to 1, which yields:

$$Z(\omega) = -\frac{V}{j\omega A {\varepsilon_{33}}^S}, \qquad Y(\omega) = \frac{1}{Z(\omega)} \qquad (20)$$

For the single layer X-cut $LiNbO_3$ case with a thickness of 300 nm and an area of 400 $\mu m^2$, we can plot the admittance response visible in **Fig. S2c**. We observe the main $SH_{1,slow}$ resonance and the in-band $SH_{1,fast}$ spurious as in **Fig. S1b**.

The resonance is observed in the electrical response, but we can also confirm the presence of the mechanical resonance in this system, by solving:

$$\boldsymbol{A}\,\boldsymbol{a} = 0 \qquad (21)$$

Since the determinant method is numerically unstable and provides no modal information, we instead use singular value decomposition (SVD), which yields

$$\boldsymbol{A} = \boldsymbol{U\Sigma V^H} \qquad (22)$$

$\boldsymbol{\Sigma}$ is a diagonal matrix containing the singular values σ of the system and the number of non-zero σ defines the rank of the matrix. When the minimum singular value $\sigma_{min}(A(\omega))$approaches zero, the system approaches a mechanical resonance. Plotting $1/\sigma_{min}$therefore reveals peaks corresponding to the antiresonance frequencies in our case since we use the stiffened coefficients. The second plot in **Fig. S2c** clearly exhibits 2 peaks corresponding to the antiresonance frequencies of the two shear modes.

To link mechanical resonance and electrical excitation, we go back to (13) where the vector **b** represents the electrical forcing. The matrix **U** from the SVD decomposition corresponds to left singular vectors in the constraint space where the vector $\mathbf{u_{min}}$ associated to $\sigma_{min}$ represents a specific pattern of boundary conditions. For a mechanical mode to be electrically excitable, its boundary traction pattern must have a non-zero projection onto the traction distribution imposed by the electrical forcing. We introduce the normalized projection representing this condition :

$$\tilde{a} = \frac{|{u_{min}}^T b|}{||b||} \qquad (23)$$

Applying this to the single X-Cut $LiNbO_3$ case, we observe a non-zero projection across the whole frequency range but with a step in between the modes (**Fig. S2c**). This projection is directly linked to the coupling of a mode and in this case the antiresonance of the main mode is more strongly coupled compared to the spurious mode. This matches very well the admittance response.

The full analysis enables to observe the admittance response, mechanical resonances and electrical excitability criteria over any frequency range using only a one dimensional wave equations analysis. This framework allows distinguishing between mechanically existing modes and those that can be electrically excited

**Bilayer analysis**

To apply the analysis to the bilayer case, the configuration is slightly different and some additional constraints are needed. Indeed, having 2 layers with different crystallographic orientations (**Fig. S2b**) results in a displacement field for each slab and therefore eight unknown wave amplitudes.

The slabs 1 and 2 thicknesses are respectively $h_1$ and $h_2$ with the interface at z=0. Similarly to the single slab problem, the traction free boundaries yield 4 equations, to which we can add 4 more by imposing continuity of displacement and stress at the interface between the slabs:

$$\begin{aligned} T_4^{(1)}(-h_1) = 0, \qquad T_5^{(1)}(-h_1) = 0 \\ T_4^{(2)}(h_2) = 0, \qquad T_5^{(2)}(h_2) = 0 \end{aligned} \tag{24}$$

$$\begin{aligned} u_x^{(1)}(0) = u_x^{(2)}(0), \qquad u_y^{(1)}(0) = u_y^{(2)}(0) \\ T_4^{(1)}(0) = T_4^{(2)}(0), \qquad T_5^{(1)}(0) = T_5^{(2)}(0) \end{aligned} \tag{25}$$

We also impose continuity of $D_3$ across the interface to be able to get the impedance of the system.

Evaluating the different quantities across the same frequency range for a bilayer configuration of 300 nm X-cut $LiNbO_3$ and 300 nm of -X-cut $LiNbO_3$ rotated by -66°, we obtain the results presented in **Fig. S2d.** First, the admittance response is exhibiting a strong resonance around 6 GHz, spurious free as previously obtained in FEM simulations (example in **Fig. 2d**). The singular values analysis shows 4 different peaks corresponding to the following modes: $SH_{1,slow}$ and its in-band $SH_{1,fasst}$ mode, $SH_{2,slow}$ and its in-band $SH_{2,fast}$ mode. But as seen in the admittance response, the $SH_{1,slow}$ mode is not coupled because of the P3F working principle and only the in-band shear mode is very weakly coupled. For the main $SH_{2,slow}$ mode exhibiting strong coupling, we can see where the antiresonance of the $SH_{2,fast}$ mode would lie even though it is not electrically visible.

Finally, the projection analysis yields very interesting results, we can see that between 3.5 and 7.4 GHz where the $SH_{1,slow}$ and the $SH_{2,fast}$ shear modes resonate mechanically, the projection criteria is more than 10 orders or magnitude lower. This indicates a strong suppression of these modes, verified in the admittance response since only the main $SH_{2,slow}$ and a small peak at 3.3 GHz appear.

Further analysis of the row-by-row contribution of the projection $u_{min}$onto $b$ shows that, for the spurious mode, the traction distribution is arranged such that the electrical forcing pushes equally in opposite directions. This results in zero net work and therefore an extremely small projection $\tilde{a}$, indicating null electromechanical coupling.

As seen in **Fig. 2**, the mode suppression is not related to a specific combination of piezoelectric coefficients, since it appears for all twisting angle configurations, but rather to the traction distribution imposed by the bilayer boundary conditions.

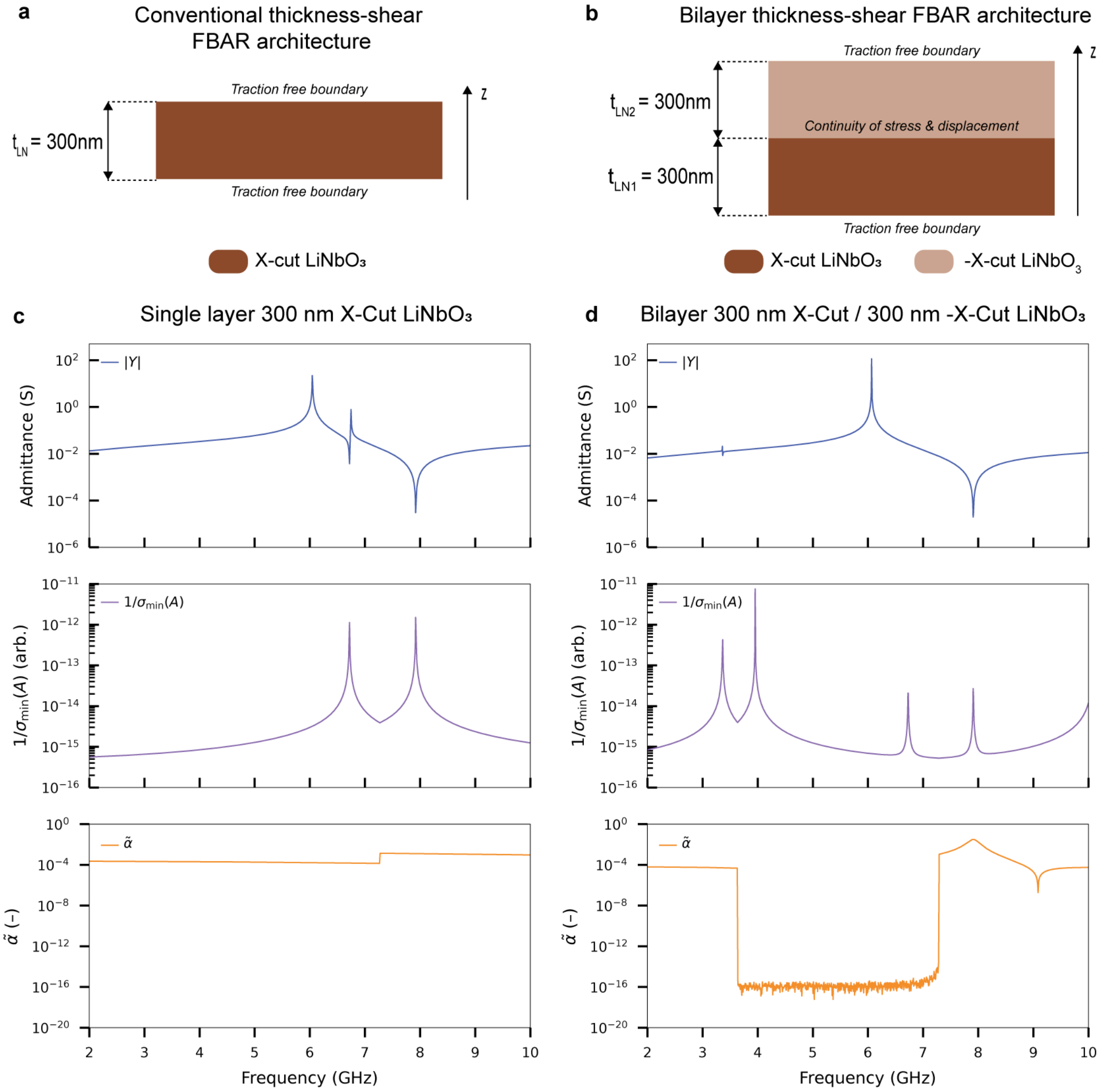

**Fig. S2 One-dimensional wave equation analysis of single-layer and bilayer $LiNbO_3$ thickness-shear FBAR resonators. a** Analytical model of a 300 nm thick single-layer X-cut $LiNbO_3$ slab with traction-free boundaries. **b** Analytical model of a bilayer composed of 300 nm X-cut $LiNbO_3$ and 300 nm −X-cut $LiNbO_3$, with traction-free outer boundaries and continuity of stress and displacement at the interface. **c, d** From top to bottom: calculated electrical admittance (top), inverse minimum singular value of the boundary condition matrix $\boldsymbol{A}$ identifying mechanical resonances (middle), and normalized projection $\tilde{\alpha}$ quantifying the electrical excitability of the modes (bottom) for the single layer **(c)** and bilayer **(d)** configurations, respectively. In the single layer structure, both slow and fast shear resonances are mechanically supported and electrically excitable. In the bilayer structure, the mechanical resonance spectrum is preserved, whereas the projection criterion approaches zero for the fast shear modes, indicating that these modes remain mechanically supported but become electrically non-excitable. Consequently, only the slow shear resonances appear prominently in the calculated admittance, reproducing the suppression of the fast shear modes observed in FEM simulations. The projection $\tilde{\alpha} = |u_{\min} b| / \| b \|$ represents the overlap between the electrically imposed traction pattern and the mechanical boundary traction associated with each resonance.

To further support this interpretation, the same analytical framework was applied to single-layer Y-cut and YX 163° $LiNbO_3$ resonators (**Fig. S3**). In both orientations, the fast shear mode remains mechanically supported, as evidenced by the singular value analysis, while its electrical excitability vanishes.

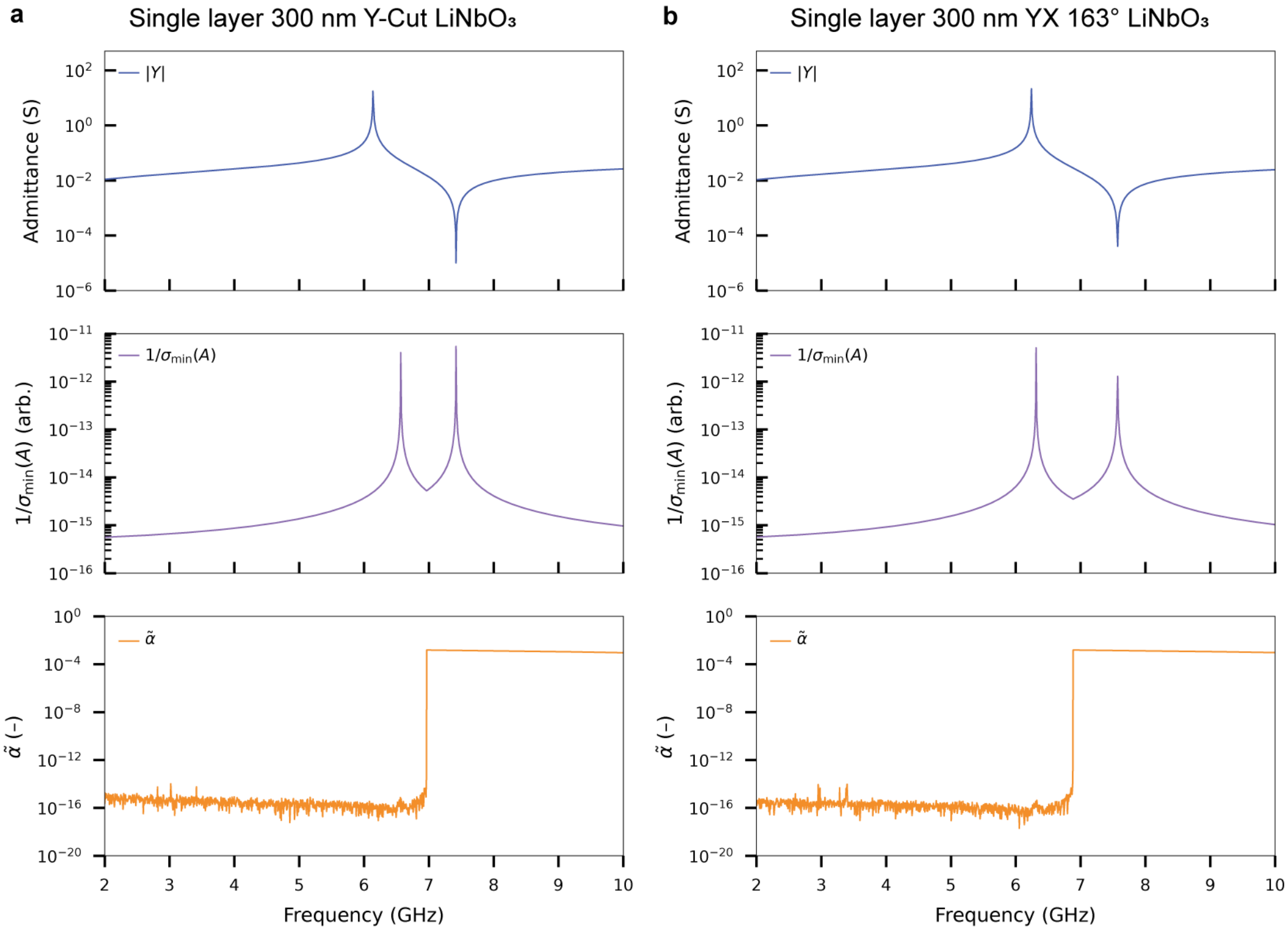

**Fig. S3 Analytical evaluation of fast shear-mode excitability for different $LiNbO_3$ orientations. a, b,** Calculated electrical admittance (top), inverse minimum singular value of the boundary-condition matrix (middle), and normalized electrical excitability projection $\tilde{\alpha}$ (bottom) for 300-nm-thick Y-cut **(a)** and YX 163° **(b)** $LiNbO_3$ resonators. In both cases, the fast shear mode remains mechanically supported but is electrically non excitable.

# Section 3 - Transverse and longitudinal spurious modes in bilayer thickness-shear FBAR configuration

A well-known limitation of thickness-shear FBAR or YBAR resonators is the presence of longitudinal and transverse spurious modes in the admittance response. In conventional designs, longitudinal spurious modes are often mitigated by etching trenches in the non-metallized regions of the resonator.

The bilayer configuration introduced in **Fig. 1** also integrates trenches as a design parameter. To evaluate the efficiency of trenches in the bilayer architecture, two cutting planes denoted A-A and B-B, were defined along the longitudinal and transverse directions , respectively, as represented in **Fig. S4a**. The corresponding two dimensional cross-sections were modelled to be used in finite-element simulations and the resulting simulated admittance responses and their real parts are shown in **Fig. S4b**. In these simulations, the resonator pitch was 15 µm, the aperture 50 µm and the trench depth 400 nm for a total piezoelectric thickness of 500 nm. Despite the use of deep trenches, spurious modes are observed in both directions, indicating a more complex spurious mode behaviour when bonding thin films with different crystallographic orientations.

Since the electromechanical coupling of the resonator is independent of the in-plane rotation of the device on the bilayer, we investigated whether a specific rotation angle could mitigate transverse and longitudinal spurious modes. **Fig. S4c** introduces the analysis method used to quantify the presence of spurious modes.  Starting from the simulated admittance response, the contribution of the fitted main mode is removed, yielding a residual response in which the prominence of in-band modes can be identified. The bilayer thickness shear FBAR resonator was rotated on the X-cut/ -66° -X-cut $LiNbO_3$ stack, and for each orientation the cumulative prominence of transverse and longitudinal spurious modes was extracted.

The angular dependence of the spurious modes prominence is shown in **Fig. S4d**. A clear 180° phase shift is observed between transverse and longitudinal responses, indicating that minimizing spurious modes in one direction will inevitably enhance them in the orthogonal direction. Similar to the electromechanical coupling, which is independent of the in-plane rotation of the resonator on the bilayer, the total spurious mode energy remains approximately conserved and redistributed between longitudinal and transverse directions. This behaviour demonstrates that trenches act as a partial filtering mechanism but cannot fully suppress spurious modes in this configuration, nor do they allow the identification of a preferential rotation angle.

Consequently, while trenches were retained in this work as a simple and effective mean to reduce spurious modes without additional optimization, further suppression techniques will be required to achieve a spurious free bandwidth. Approaches such as IDT apodization, piston-mode design or use of irregular electrodes commonly employed in FBAR technology represent promising directions for future optimization of the bilayer thickness-shear FBAR resonator response.

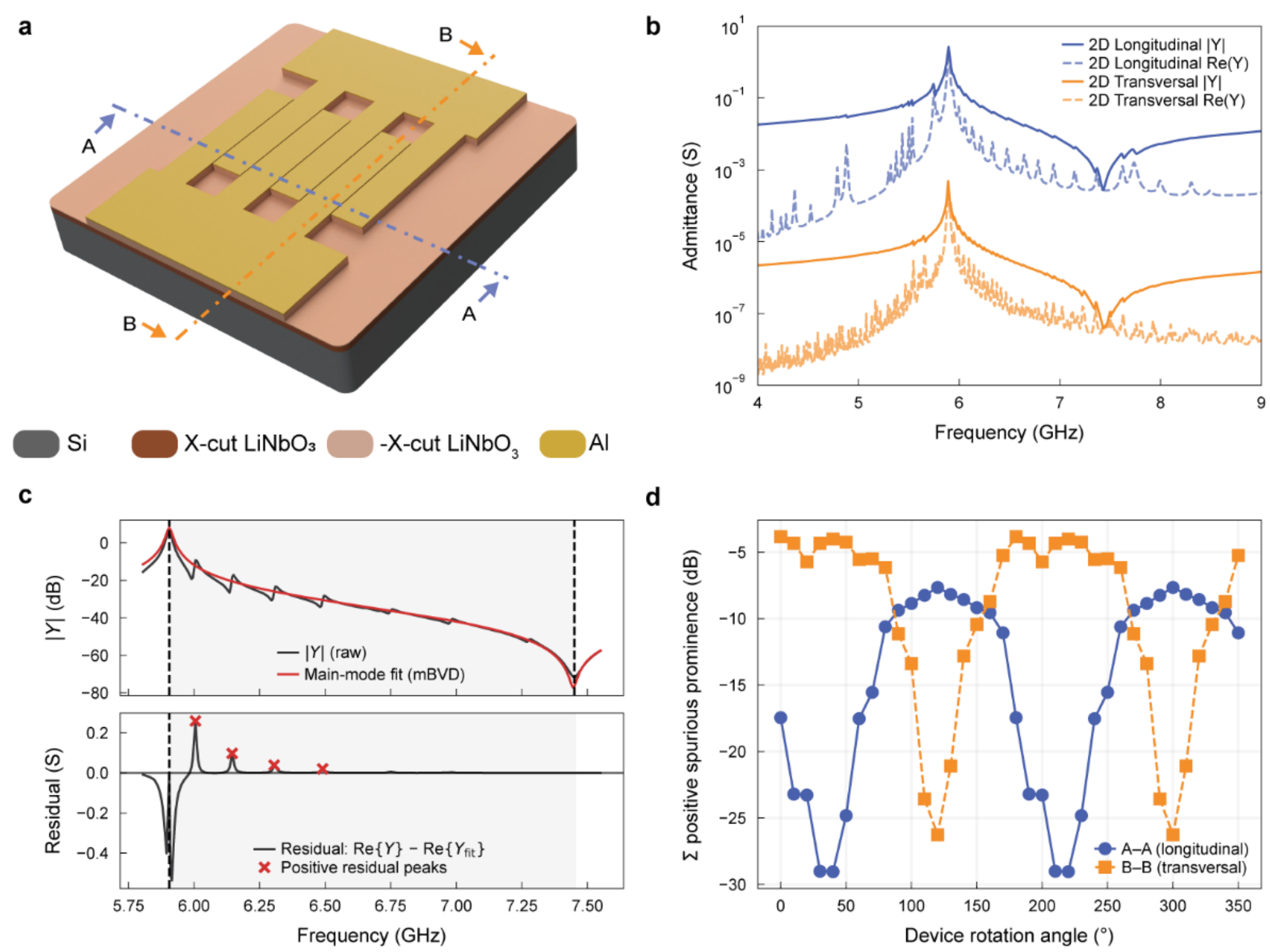

**Fig. S4 Longitudinal and transversal spurious-modes prominence under in-plane rotation. a** Top-view schematic of the bilayer thickness-shear FBAR with 2 cutting planes used for 2D simulations: A-A highlighting longitudinal spurious modes and B-B highlighting transverse spurious modes. **b** Simulated admittance responses extracted using A-A and B-B cross-sections, highlighting weak but numerous longitudinal and transverse spurious modes. Curves are offset for clarity. **c** Illustration of the spurious mode quantification method. The fitted main mode response is subtracted from the simulated admittance to obtain a residual response from which the prominence of in band spurious modes is extracted. **d** Sum of the extracted prominence of spurious modes for the longitudinal (A-A) and transverse (B-B) directions as a function of the in-plane rotation angle of the resonator, for a fixed bilayer twist angle of $\theta$ =294°. The plot reveals an angular dependence of the spurious modes prominence with a 180° phase shift, meaning that minimizing longitudinal spurious modes maximizes transverse ones, and vice versa.

# Section 4 - Influence of electrode thickness asymmetry

**Fig. S5** summarizes the influence of the electrode symmetry on shear mode excitation in bilayer thickness-shear bulk acoustic resonators. In contrast to the symmetric 75 nm/75 nm electrode configuration considered in **Fig. 3**, the simulations in **Fig. S5a** and **S5b** compare this configuration with asymmetric 25 nm/75 nm electrodes while varying the piezoelectric thickness ratio. With asymmetric electrodes, the maximum $SH_{2,slow}$ coupling and minimum $SH_{1,slow}$ coupling are shifted towards lower $t_{LN,top}/t_{LN,bot}$ ratios, reflecting the additional asymmetry introduced by the unequal electrode thicknesses.

The $SH_{2,slow}$ coupling exhibits a more pronounced maximum for the asymmetric electrode configuration, rather than the broad plateau obtained with symmetric electrodes, indicating reduced robustness to variations in the piezoelectric thickness ratio. Interestingly, for thinner top $LiNbO_3$ layers, the decrease in $SH_{2,slow}$ coupling is less pronounced with asymmetric electrodes. In parallel, the $SH_{1,slow}$ mode is more strongly excited for thicker top layers in the asymmetric configuration.

These results highlight the crucial role of electrode symmetry in the bilayer architecture. Taken together with the piezoelectric thickness study of **Fig. 3**, these findings demonstrate that both $LiNbO_3$ layer thickness balance and electrode symmetry are important design parameters for maintaining high $SH_{2,slow}$ coupling while suppressing the $SH_{1,slow}$ mode. Electrode thickness therefore provides an additional degree of freedom for controlling the modal balance and resonance frequency.

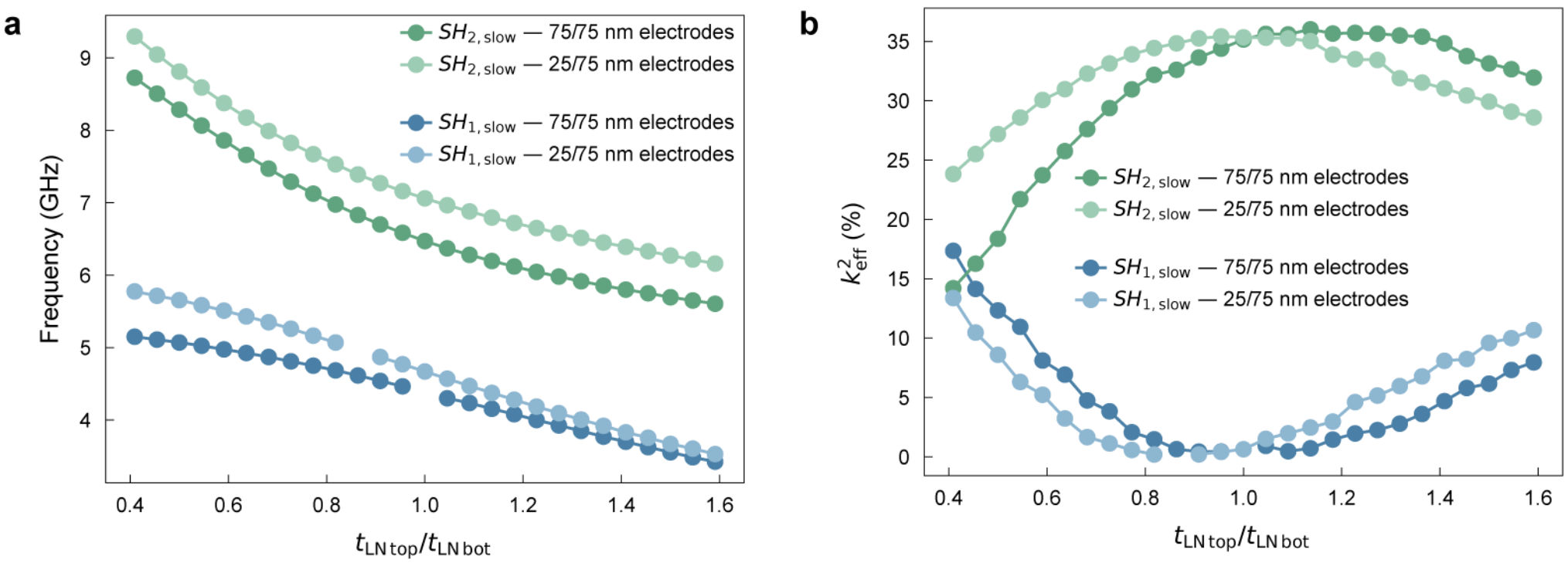

**Fig. S5 | Influence of electrode symmetry on shear-mode excitation in bilayer thickness-shear bulk acoustic resonators. a** Simulated resonance frequencies of the $SH_{1,slow}$ and $SH_{2,slow}$ modes as a function of the $LiNbO_3$ thickness ratio $t_{\mathrm{LN,top}}/t_{\mathrm{LN,bot}}$, comparing symmetric (75 nm/75 nm) and asymmetric (25 nm/75 nm) electrode configurations. **b** Corresponding simulated electromechanical coupling factors $k^2_{\mathrm{eff}}$. Electrode asymmetry shifts the thickness ratio corresponding to maximum $SH_{2,slow}$ coupling and minimum $SH_{1,slow}$ coupling and modifies the robustness of the modal response to $LiNbO_3$ thickness imbalance.

# Section 5 - Third and fifth-order ladder-type filter measurements results

Third and fifth-order ladder-type filters were fabricated using bilayer resonators with 220 nm/215 nm and 220 nm/365 nm X-cut/-X-cut $LiNbO_3$ thickness combinations to achieve the required frequency spacing between the series and shunt resonators. The broadband synthesized and measured responses are displayed in **Fig. S6**. The third-order filter synthesized and measured responses are shown in **Fig. S6a** and **S6b**, while the fifth-order responses, previously discussed in the main text (**Fig.4)** are presented in **Fig. S6c** and **S6d** to allow comparison over a wider frequency range.

Overall, the measured responses agree well with the synthesized filters obtained from finite-element simulations and mBVD-based circuit synthesis. As discussed in the main text, the main discrepancy is the shift in centre frequency caused by fabrication-induced thickness deviations, which also modifies the achieved fractional bandwidth. The higher insertion loss observed experimentally mainly originates from the lower resonator quality factors, despite the inclusion of additional series resistive losses ($R_s$ = 1.9 Ω) during the filter synthesis.

The broadband responses also illustrate a limitation of achieving the required frequency spacing through unbalanced $LiNbO_3$ layer thicknesses. Since both resonator types are fabricated on the same chip, only the top $LiNbO_3$ layer can be trimmed after bonding. Increasing the resonance frequency by reducing the top layer thickness simultaneously increases the asymmetry of the bilayer, moving the structure away from the symmetry condition responsible for the destructive interference of the $SH_{1,slow}$ and $SH_{3,slow}$ modes. As a result, the coupling of these modes increases while the coupling of the desired $SH_{2,slow}$ mode is no longer maximized. This leads to additional transmission bands below and above the main filter passband, as observed in the broadband responses.

Consequently, the bilayer thicknesses must be selected by simultaneously considering the target centre frequency, fractional bandwidth, and modal coupling. Although this introduces an additional design constraint compared to conventional single-layer resonators, appropriate thickness combinations still enable large FBW while maintaining sufficient separation from neighbouring passbands.

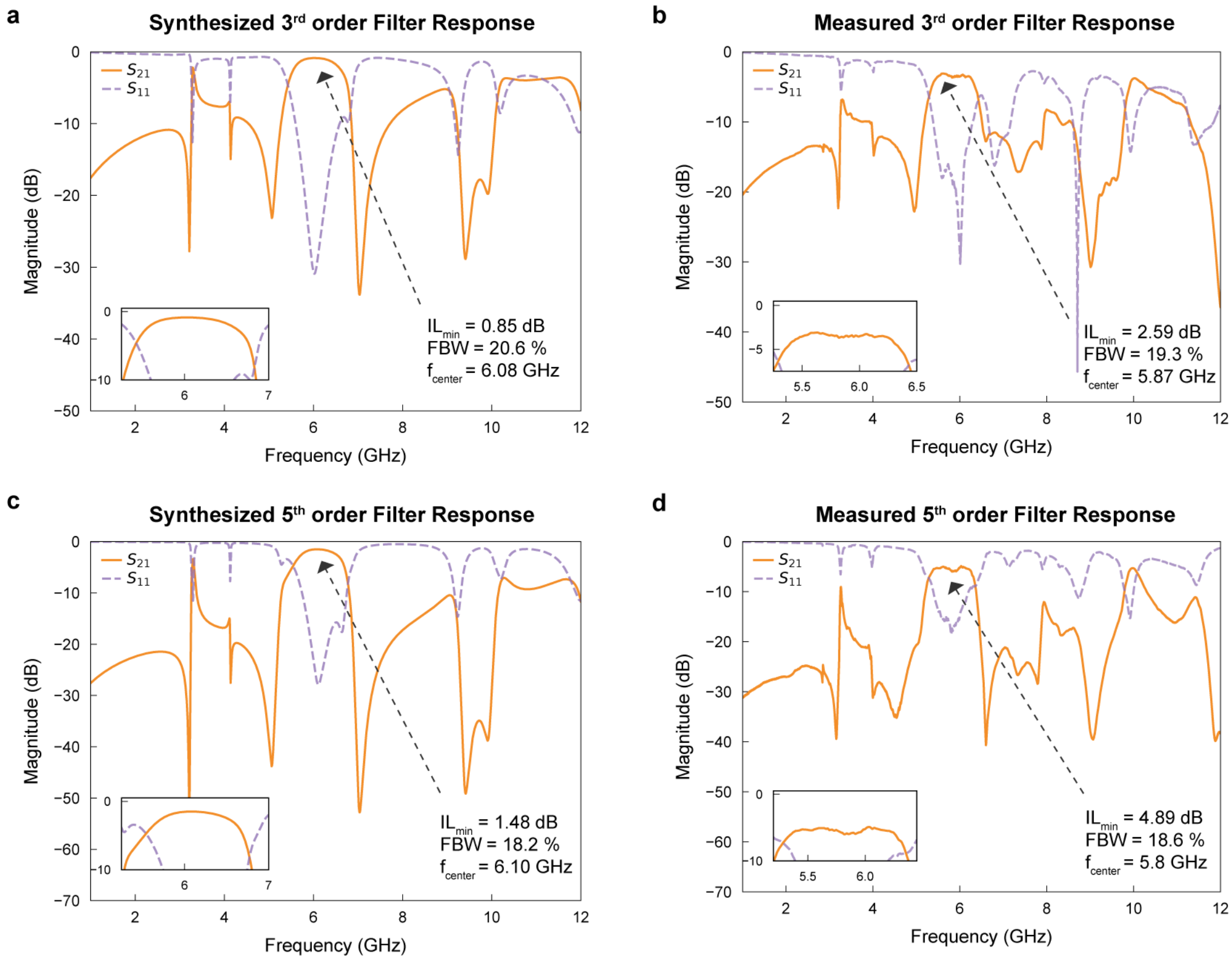

**Fig. S6 Synthesized and measured responses of 3rd and 5th order ladder-type filters based on bilayer $LiNbO_3$ thickness-shear FBAR resonators. a** Synthesized response of the third-order ladder filter obtained from finite-element simulations and mBVD-based circuit synthesis. **b** Measured response of the fabricated third-order ladder filter. **c** Synthesized response of the fifth-order ladder filter. **d** Measured response of the fabricated fifth-order ladder filter. Insets show the filter passbands together with the extracted insertion loss, centre frequency, and fractional bandwidth.